\documentclass[twocolumn,secnumarabic,nobibnotes,superscriptaddress,aps,nofootinbib,10pt,pra]{revtex4-2}
\usepackage{color, xcolor, colortbl}
\usepackage{graphicx}
\usepackage{amsmath,amssymb,amsthm,amsfonts}
\usepackage{mathtools}
\usepackage[ruled,vlined,linesnumbered]{algorithm2e}
\usepackage{algpseudocode}
\usepackage{dcolumn}
\usepackage{epstopdf}
\usepackage{bm}
\usepackage{appendix}
\usepackage{multirow}
\usepackage{braket}
\usepackage[english]{babel}
\usepackage[T1]{fontenc}
\usepackage{xpatch}
\usepackage{adjustbox}
\usepackage{xspace}
\usepackage[roman]{complexity}
\usepackage{xparse}
\usepackage{comment}
\usepackage{balance}
\usepackage{booktabs}
\usepackage[table]{xcolor}
\usepackage{adjustbox}
\usepackage{hyperref}
\newcommand{\ptwo}[2]{\( #1 \otimes #2 \)}
\newcommand{\hlptwo}[2]{\cellcolor{red!12}\( #1 \otimes #2 \)}

\newcommand{\MarylandMath}{Department of Mathematics, University of Maryland, College Park, MD 20742, USA}
\newcommand{\MarylandCS}{Department of Computer Science, University of Maryland, College Park, MD 20742, USA}
\newcommand{\MarylandQuICS}{Joint Center for Quantum Information and Computer Science, University of Maryland, College Park, MD 20742, USA}
\newcommand{\ArgonneMathCS}{Mathematics and Computer Science Division, Argonne National Laboratory, Lemont, IL 60439, USA}

\begin{document}
\title{Randomized Subsystem Descent for Fermion-to-Qubit Mapping}
\author{Gengzhi Yang}
\affiliation{\MarylandMath}
\affiliation{\MarylandQuICS}

\author{Di Wu}
\affiliation{\MarylandMath}

\author{Haizhao Yang}
\thanks{hzyang@umd.edu}
\affiliation{\MarylandMath}
\affiliation{\MarylandCS}

\author{Xiaodi Wu}
\thanks{xwu@cs.umd.edu}
\affiliation{\MarylandQuICS}
\affiliation{\MarylandCS}

\author{Ji Liu}
\thanks{ji.liu@anl.gov}
\affiliation{\ArgonneMathCS}

\date{Latest revision: \today}

\begin{abstract}
We propose a versatile and efficient algorithmic framework for optimizing fermion-to-qubit mappings by generalizing the idea of randomized block coordinate descent. Our greedy approach, termed Randomized Subsystem Descent, iteratively samples a tractable subsystem from the full Hamiltonian, performs optimization within the subsystem under a given metric, and then reintegrates the updated subsystem into the global operator. Restricting the optimization to a subsystem at each iteration ensures computational efficiency, bypassing the dimensional bottlenecks that usually hinder global search heuristics. We benchmark our algorithm on one- and two-dimensional lattice hopping models, the Hubbard model with up to $16 \times 16$ sites, alongside a collection of molecular electronic-structure Hamiltonians with up to $54$ modes and more than $180,000$ Pauli strings. Across all benchmarks, our method consistently provides appreciable reduction in (weighted) Pauli weight, suggesting that Randomized Subsystem Descent is a practical and scalable framework for lowering the resource overhead of finding hardware-efficient Hamiltonian encodings.
\end{abstract}
\maketitle

\section{Introduction}
Simulating fermionic systems is widely viewed as one of the most promising applications of quantum computers~\cite{feynman2018simulating,wecker2015solving,barkoutsos2017fermionic,lanyon2010towards}.
To execute such simulations on quantum platforms, one must first encode fermionic operators into qubit operators via fermion-to-qubit mappings.
The choice of fermion-to-qubit mapping would directly impact the cost of Hamiltonian simulation, since different mappings produce qubit Hamiltonians with different Pauli weights and locality structures.

The design of efficient fermion-to-qubit mappings has attracted significant attention in recent years~\cite{chiew2023discovering,liu2024fermihedral,yu2025clifford,liu2025hatt,setia2019superfast,derby2021compact}.
Beyond several other established methods~\cite{jordan1928paulische,bravyi2002fermionic,tranter2018comparison,jiang2020optimal,verstraete2005mapping,derby2021compact2,o2024ultrafast}, optimization-based fermion-to-qubit mappings~\cite{steudtner2019quantum, nys2022variational, miller2026treespilation, liu2024fermihedral} are favored in practice because they can be adapted to different circuit compilers~\cite{li2022paulihedral,de2020architecture,gui2020term,cowtan2019phase,van2020circuit} through problem-dependent objective functions. A recent work~\cite{yu2025clifford} introduced a novel framework that formulates fermion-to-qubit mapping as a Clifford-circuit optimization problem. By employing simulated annealing to minimize the Pauli weight of the target Hamiltonian, this approach achieves strong empirical performance. However, this global-search strategy can be computationally expensive in practice, with optimization runs reported to take up to 3 days on a single CPU. 

\begin{figure}
    \centering
    \includegraphics[width=0.95\linewidth]{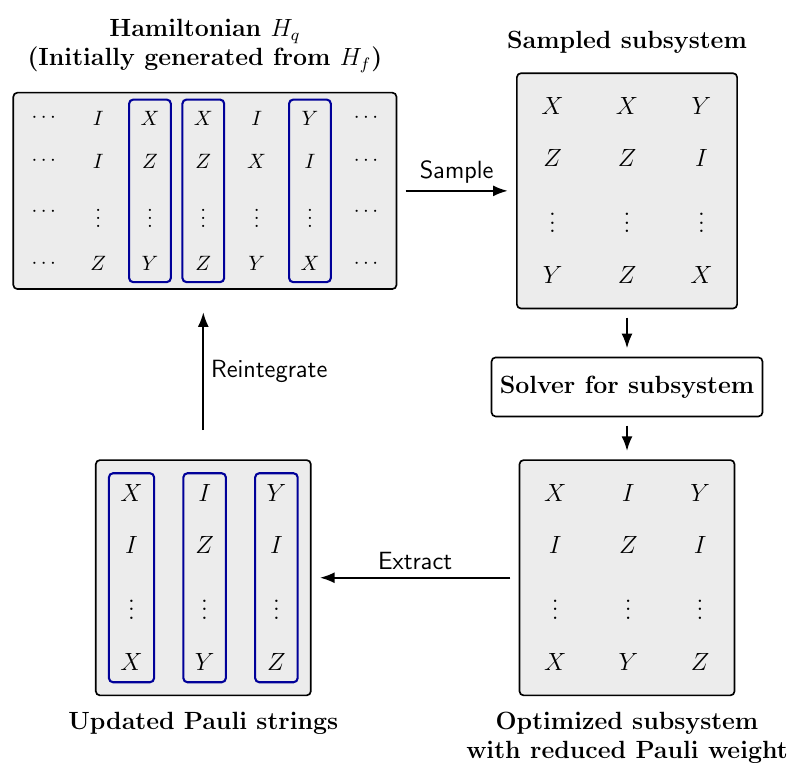}
    \caption{Workflow for the Randomized Subsystem Descent.}
    \label{fig:schematic}
\end{figure}

In this paper, we attempt to develop a more efficient optimization-based framework for generating fermion-to-qubit mappings for large systems. To this end, we introduce Randomized Subsystem Descent (RSD), a simple yet efficient heuristic for designing fermion-to-qubit mappings and problem-aware Hamiltonian encodings.
Our method is inspired by Block Coordinate Descent (BCD)~\cite{xu2013block,nesterov2012efficiency,lu2015complexity,necoara2017random}, a classical optimization technique in which a high-dimensional problem is decomposed into smaller subproblems that are solved sequentially.
Similar ideas also appear in peephole optimization~\cite{prasad2006data} for Clifford circuit synthesis, where the previous work~\cite{bravyi2021clifford} used symbolic peephole optimization that repeatedly projects a Clifford circuit onto small qubit subsets.
We adapt this local-optimization philosophy to the discrete and non-smooth optimization landscape of Pauli-weight minimization. 
As visualized in Figure~\ref{fig:schematic},
at each iteration, a small subset that contains $k$ qubits is sampled from the full $n$-qubit Hamiltonian, and a subsystem solver is applied to minimize the objective function, e.g. the Pauli weight, within that subsystem.
The optimized subsystem is then reintegrated back into the global Hamiltonian. 
A key feature of the algorithm is that it only accepts updates that reduce the total Pauli weight, making it a greedy method.
Despite the lack of theoretical guarantees of global optimality, this strategy is highly effective in practice when combined with an appropriate sampling method.

We first benchmark RSD on the one- and two-dimensional lattice hopping models. 
For the 1D chain hopping model, across all random seeds, RSD consistently provides qubit Hamiltonians with a lower average Pauli weight than conventional mappers like Jordan-Wigner, Bravyi-Kitaev, and the ternary tree mapping. The percentage reduction against conventional methods is substantial:
In models with varying hopping ranges, improvements fluctuate but consistently stabilize near 10\% for longer ranges.
In all-to-all connection models, the reduction ranges from over 30\% for small systems down to approximately 8\% to 10\% as the system size scales to $20$.  Our method achieves these reductions while matching or exceeding the performance of H+CNOT annealing. Even more, for a specific system with a known global optimum, RSD successfully converged to the exact solution, whereas simulated annealing failed to do so.
For the 2D nearest-neighbor hopping models, 
RSD yields an over 40\% Pauli weight reduction for grid side lengths smaller than 10, and around 20\% improvement for larger system sizes.

We further evaluate RSD on the Hubbard model to assess its scalability and effectiveness beyond hopping-only lattice models.
For the Hubbard model evaluated on grid side lengths up to $16$, our method scales to system sizes well beyond the reach of simulated annealing. Furthermore, it achieves substantial Pauli-weight reductions over conventional mappers, peaking at over 30\% for intermediate system sizes. Even at the $16\times 16$ grid, RSD still managed to provide a more than 10\% reduction in Pauli weight compared with the conventional mappers.

To evaluate performance on molecular electronic-structure Hamiltonians, which provide a more realistic setting for quantum chemistry, we tested our method using both STO-3G and 6-31G basis sets across a diverse set of systems under two different metrics, namely the total Pauli weight and weighted Pauli weight. For all tested molecules and basis sets, RSD achieves the best performance on both metrics.
Notably, this includes reductions in total Pauli weight of nearly 20\% for the $\mathrm{H}_2$ molecule in the STO-3G basis and more than 10\% for the LiH and O$_2$ molecule in the 6-31G basis.
For the weighted Pauli weight, our approach yields substantial reductions, peaking at approximately 25\% for NaF and $\text{CO}_2$ in the STO-3G basis and $\text{LiH}$, $\text{O}_2$ and $\text{NaF}$ in the 6-31G basis.

Our method finishes within hours for systems tested, allowing one to deal with larger systems where previous methods struggle. This suggests that the fermion-to-qubit mapping problem is well suited to greedy local optimization and may not require global-search heuristics. More broadly, the RSD framework is not limited to fermion-to-qubit mapping only, but can also be viewed as a general approach for constructing efficient Hamiltonian encodings, thereby extending the scope of previous work on Hamiltonian embedding~\cite{leng2025expanding, li2026resource}.

The rest of this paper is organized as follows:
Section~\ref{sec:method} presents the technical background on fermion-to-qubit mappings, introduces the Randomized Subsystem Descent algorithm, and describes the subsystem Clifford solver along with different sampling strategies.
Section~\ref{sec:results} reports the numerical results, including the lattice models (\ref{sec:lattice}), the Hubbard model (\ref{sec:hubbard}) and molecular systems (\ref{sec:mols}).
Section~\ref{sec:conclusions} concludes with a discussion of implications and future directions.

\section{Methods}

\label{sec:method}

\subsection{Background}
Quantum simulation constitutes a primary application of quantum computing.
Recent algorithmic advances on Hamiltonian simulation~\cite{gilyen2019quantum,an2021time,berry2014exponential,berry2015simulating,kieferova2019simulating,childs2021theory} have revealed the great potential of simulating fermionic systems on quantum computers.
Consequently, the design of efficient fermion-to-qubit mappings is crucial for accurately representing fermionic interactions on quantum hardware. Say a fermionic system is constructed by $n$ fermionic modes, then it can be described by the creation and annihilation operators $\{a_i^\dag\}_{i=0}^{n-1}$ and $\{a_i\}_{i=0}^{n-1}$, where they obey the canonical anticommutation relations~\cite{pauli1925zusammenhang}
\begin{equation}
\begin{split}
    \{a_i, a_j\} = \{a_i^\dag, a_j^\dag\} &= 0,\\
    \{a_i^\dag, a_j\} &= \delta_{ij}I,
\end{split}
\end{equation}
where $\delta_{ij}$ is the Kronecker delta and $I$ is representing the identity.

Alternatively, one can use the Majorana operators 
\begin{equation}
    m_{2i} := a_i^\dag + a_{i}, \quad m_{2i+1} := \imath (a_i^\dag - a_i)
\end{equation}
as the generators of the Fermionic system. The Majorana operators satisfy another anticommutation relation as $\{m_i, m_j\} = 2\delta_{ij}I$. Due to the simplicity of this relation, people usually consider constructing mutually anticommuting Pauli operators~\cite{jiang2020optimal} to represent the Majorana operators.

While general fermion-to-qubit mappings do not strictly require an equivalence between the number of fermionic modes and the number of qubits~\cite{nys2022variational,setia2019superfast,steudtner2019quantum}, we focus on the cases where these two dimensions are equal. Nevertheless, we emphasize that the proposed methodology is highly adaptable and can be readily extended to accommodate more general mappings.
Consider a $2^n$-dimensional fermionic Fock space $\mathcal{H}_f$ and an $n$-qubit Hilbert space $\mathcal{H}_q$,
a fermion-to-qubit mapping can be defined as an isometry $\phi: \mathcal{H}_f \to \mathcal{H}_q$ where
\begin{equation}
    \phi: m_{i} \mapsto P_i.
\end{equation}
Here $P_i \in \{I,X,Y,Z\}^{\otimes n}$ and they satisfy
\begin{equation}
    \{P_i, P_j\} = 2\delta_{ij}I.
\end{equation}
One of the simplest conventional mapping is the Jordan-Wigner mapping~\cite{jordan1928paulische}, where the Majorana operators are mapped into
\begin{equation}
    m_{2i} \rightarrow \left(\prod_{k=0}^{i-1} Z_k\right)X_i, \quad
    m_{2i+1} \rightarrow \left(\prod_{k=0}^{i-1} Z_k\right)Y_i.
\end{equation}

The primary goal of optimizing fermion-to-qubit mappings is to enable more efficient Hamiltonian simulation, the cost of which is highly related to the Pauli weight of the mapped Hamiltonian in the qubit space. Consider an operator $O$ acting on $n$ qubits expressed as a linear combination of Pauli strings
\begin{equation}
    O = \sum_j c_j P_j
\end{equation}
where $c_j \in \mathbb{R}$ and $P_j \in \{I, X, Y, Z\}^{\otimes n}$ is a tensor product of $n$ Pauli matrices. The Pauli weight of this operator, denoted as $\text{PW}(O)$, is defined as the sum of the weights of its constituent strings
\begin{equation}
    \text{PW}(O) := \sum_j \text{wt}(P_j).
\end{equation}
We use $\text{wt}(P_j)$ to denote the Hamming weight of the Pauli string $P_j$, defined as the number of non-identity tensor factors within $P_j$. This metric is closely related to the cost of the Trotter formulae~\cite{childs2021theory}, where the first-order approximation is
\begin{equation}
    e^{-\imath O t} \approx \left(\prod_j e^{-\imath P_j c_j t/M}\right)^M.
\end{equation}
Here $t$ is the simulation time and $M$ is the number of time steps.
Minimizing the total Pauli weight can be viewed as a procedure of reducing the cost if we directly apply these operator-splitting methods~\cite{aftab2024multi,an2022time}.

Other choices of Hamiltonian simulation algorithms would lead to different metrics to judge fermion-to-qubit mappings. For example, for the qDRIFT algorithm~\cite{campbell2018random}, a more appropriate objective should be the weighted Pauli weight~\cite{li2025huffman} as
\begin{equation}
    \text{wPW}(H) := \sum_j |c_j| \text{wt}(P_j).
\end{equation}

\subsection{Optimization of fermion-to-qubit mappings via unitary transformations}
Given two distinct fermion-to-qubit mappings $\phi_1$ and $\phi_2$, some fermionic operator $O_F$ acting on $\mathcal{H}_f$ can be mapped to the corresponding qubit operators $\phi_1 O_F \phi_1^\dagger$ and $\phi_2 O_F \phi_2^\dagger$ acting on $\mathcal{H}_q$. It follows that the first operator can be transformed into the second via the relation $U \phi_1 O_F \phi_1^\dagger U^\dagger = \phi_2 O_F \phi_2^\dagger$, where $U$ is a unitary operator acting on $\mathcal{H}_q$ such that $U\phi_1 = \phi_2$.
This algebraic relationship motivates the optimization of a given initial mapping $\phi$ by applying unitary transformations to $\phi O_F \phi^\dagger$ to identify more desirable mappings, i.e. those potentially exhibiting enhanced hardware efficiency.

Our objective is to apply transformations that preserve this linear algebraic structure while optimizing the properties of the resulting operator.
For any cost function $\text{cost}(\cdot)$ adopted as our optimization objective, the optimization problem is formulated as
\begin{equation}\label{eqn:optimization-obj}
    \underset{U \in \mathbb{U}(2^n)}{\arg\min} \; \text{cost}\left(U^\dagger O U\right),
\end{equation}
where $\mathbb{U}(2^n)$ is the unitary group of degree $2^n$, and $O$ is the observable in $\mathcal{H}_q$ obtained from an initial mapping.
In this framework, observables other than the original Hamiltonian must be transformed consistently with the Hamiltonian to evaluate their expectation values in the optimized encoding.

The selection of the objective function can be tailored to suit specific computational goals~\cite{miller2026treespilation,de2025optimised}. We select the total Pauli weight and weighted Pauli weight as our main objectives because they are closely connected the cost of Hamiltonian simulation~\cite{mukhopadhyay2023synthesizing, campbell2018random}.

\subsection{Randomized Subsystem Descent}
\label{sec:rsd}
The optimization of large-scale quantum systems frequently encounters the curse of dimensionality.
Optimizing over the entire unitary group $\mathbb{U}(2^n)$ in Equation~\ref{eqn:optimization-obj} is computationally intractable due to several reasons: one cannot apply the matrix multiplications $U^\dag O U$ easily on a classical computer, and the Pauli representation of $U^\dag O U$ may have exponentially many terms, making it even impossible to evaluate the objective function.
Furthermore, recent work~\cite{yu2025clifford} indicates that standard heuristic approaches, such as simulated annealing, fail to converge to satisfactory solutions when the number of fermionic modes exceeds 100. These profound scaling bottlenecks necessitate the development of more efficient optimization strategies.

In the multi-variable optimization, for $\bm{x} = (x_0, x_1, \cdots, x_n)$, the optimization of a function $f(\bm{x})$ can usually be decomposed to the optimization of smaller subspaces. The well-known coordinate descent method~\cite{d1959convex,xu2013block,xu2013block,shi2016primer,wright2015coordinate} is an example which successively minimizes along coordinate directions to find the minimum. Starting with $\bm{x}^{(0)} = (x_0^{(0)},x_1^{(0)}, \cdots, x_n^{(0)})$, at iteration $t$, one can sequentially do line search along individual coordinates as
\begin{equation}
\begin{split}
    &x_i^{(t+1)} =\\  
    &\underset{y}{\arg\min}\; f(x_0^{(t+1)}, \cdots, x_{i-1}^{(t+1)},y, x_{i+1}^{(t)}, \cdots, x_n^{(t)}).
\end{split}
\end{equation}
The most obvious consequence is that $f(\bm{x}^{(t+s)}) \leq f(\bm{x}^{(t)})$ for any $s \in \mathbb{N}_+$, thus one can expect a monotonically decreasing behavior. It is worth noting that coordinate descent methods have a wide-range of successful applications, including the leading eigenvalue problem~\cite{lei2016coordinate,li2019coordinatewise}, image-processing~\cite{han2018permuted}, control theory~\cite{ross2023derivation}, communication problems~\cite{wu2016fpga,wu2017high} and full configuration interaction~\cite{wang2019coordinate,zhang2025parallel}.
In fact, one can extend the idea of updating one coordinate at one time into updating a subset of the coordinates. This principle is formalized as Block Coordinate Descent (BCD)~\cite{bertsekas1997nonlinear}, where a high-dimensional problem is decomposed into smaller subsets that are optimized sequentially. Randomness is introduced for selecting the sub-blocks~\cite{nesterov2012efficiency,yuan2020block,patrascu2015random,liu2014blockwise}, underscoring the importance of effective sampling strategies. 
A related philosophy has also appeared in Clifford circuit synthesis through symbolic peephole optimization~\cite{bravyi2021clifford},
where one repeatedly isolates a small subset of qubits for optimization purposes.

To address the optimization problem of finding efficient fermion-to-qubit mappings, we propose the Randomized Subsystem Descent (RSD) in Algorithm~\ref{algo:rsd}. RSD is an algorithm that adapts the philosophy of BCD into a stochastic, greedy search over discrete operator spaces. Instead of differentiating a continuous parameter space, RSD iteratively refines an $n$-qubit Hamiltonian by targeting a manageable $k$-qubit subsystem where $k \ll n$. For simplicity, we denote the size of the subsystem as the width of RSD.

For each sampled subsystem, it only contains up to a few qubits and thus can be efficiently solved. To this end, one can always apply brute-force search for a reasonably small subsystem.
One possible approach is to use Depth-First search algorithm to deterministically explore all possible paths of applying the Clifford gates up to a given depth, as introduced in Section~\ref{sec:sub-clifford}. 
The subsystem solver can be flexible in the sense that the primary objective of the search is to identify a unitary operation that minimizes a designated metric.

\begin{algorithm}[ht!] 
\caption{Randomized Subsystem Descent}
\label{algo:rsd}
\SetAlgoLined
\KwIn{Original $n$-qubit Hamiltonian $H^{(0)}$; Cost function $c$; Solver that minimizes the cost for $k$-qubit systems; Number of iterations $T$; Sampling strategy $s$ that maps a Hamiltonian to a distribution $\mathcal{S}$ over subsets of qubit indices of size $k$.}
\KwOut{Updated $n$-qubit Hamiltonian $H^{(T)}$.}
 
\BlankLine
\For{$t = 1, 2, \dots, T$}{
    Draw a subset of qubit indices $\mathcal{I}_t \subset \{1, 2, \dots, n\}$ such that $\left|\mathcal{I}_t\right| = k$ according to distribution $\mathcal{S}^{(t)} = s\left(H^{(t-1)}\right)$\;
    Obtain a $k$-qubit unitary $V^{(t)}$ by executing the subsystem solver on the subsystem indexed by $\mathcal{I}_t$ of $H^{(t-1)}$\;
    $U^{(t)} \gets V^{(t)} \otimes I_{\mathcal{I}_t^c}$, where $I_{\mathcal{I}_t^c}$ is the identity on the unselected $(n-k)$ qubits\;
    \eIf{$c\left(\left(U^{(t)}\right)^\dag H^{(t-1)} U^{(t)}\right) < c\left(H^{(t-1)}\right)$}{
        $H^{(t)} \gets \left(U^{(t)}\right)^\dag H^{(t-1)} U^{(t)}$\;
    }{
        $H^{(t)} \gets H^{(t-1)}$; 
    }
}
\Return $H^{(T)}$
\end{algorithm}

The efficacy of the RSD algorithm hinges critically on the sampling strategy used to identify the target subsystem during each iteration.
Subsystem sampling can be approached through several distinct strategies. The simplest method is uniform sampling, which entails the random selection of $k$ qubit indices from a total of $n$ possibilities. While computationally trivial, this method is fundamentally inefficient as it remains agnostic to the current state of the whole system.
To introduce state-awareness, an alternative approach calculates the Hamming weight $h_i$ for each individual qubit index $i$. The subsystem is then sampled using a weighted probability distribution
\begin{equation}
    P_i = \frac{h_i}{\sum_{j} h_j}.
\end{equation}
Building upon this, one can exploit the fact that potential Pauli weight reduction is an inherent property of the Hamiltonian. This allows for the training of a machine learning (ML) model designed to predict the specific subsystem indices that maximize Pauli reductions.

Although the uniform sampling strategy offers the simplest implementation, it is neither fast nor computationally efficient in terms of finding desired results. The Hamming weight sampling approach utilizes the information of the present operator thus usually yields better results given a sufficiently large number of iterations.
We use the Hamming-weight sampling strategy in our numerical experiments because of its better performance and simpler implementation.
For a comparison of these three methods, one can refer to Appendix~\ref{append:sampling}.

\subsection{Subsystem solvers via Clifford transformations}
\label{sec:sub-clifford}

Building upon prior work~\cite{yu2025clifford}, we employ Clifford transformations to construct the subsystem solver. For a $k$-qubit subsystem, the Clifford gates are
\begin{equation}
    S = \{H_i, S_i, \text{CNOT}_{ij} \mid i \neq j, 1 \leq i,j \leq k\}.
\end{equation}
The unitary transformation $U$ we are seeking can thus be constructed as a finite product of these generators as
\begin{equation}
    U = \prod_j V_j,
\end{equation}
where each $V_j \in S$. 

While this restricted construction does not span the full unitary space since Clifford gates do not constitute a universal gate set, restricting the optimization to stabilizer-preserving transformations maintains computational tractability, because
the Clifford transformations would preserve the number of Pauli strings in the Hamiltonian, and it is easy to update the Hamiltonian since one does not need to actually do matrix multiplications. Appendix~\ref{append:clifford} provides a complete table of Clifford transformations acting on Pauli matrices.
It is worth emphasizing that even though searching within the Clifford group is sufficient to yield substantial improvements over conventional mapping strategies, one can use alternative subsystem solvers within the RSD framework for problem-aware fermion-to-qubit mapping design.

\section{Results}
\label{sec:results}
\begin{figure*}
    \centering
    \includegraphics[width=0.75\linewidth]{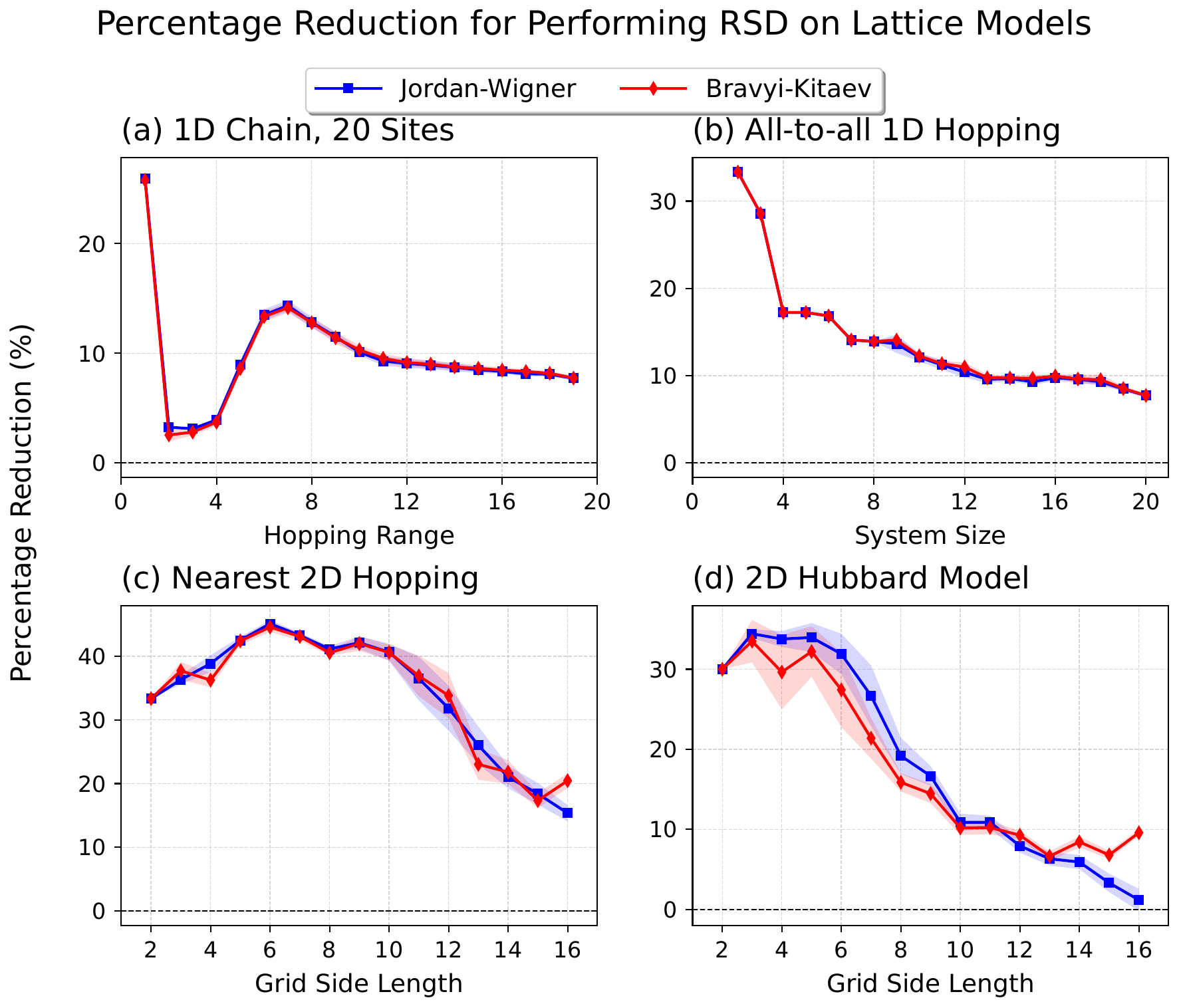}
    \caption{The percentage reduction compared with all three conventional mappers. Points show the mean over 100 random seeds, and shaded regions indicate $\pm$ one standard deviation. Blue and red denote RSD initialized with the Jordan–Wigner and Bravyi–Kitaev mappings, respectively.}
    \label{fig:latticePMstd}
\end{figure*}

The subsequent sections detail our numerical results for one- and two-dimensional hopping models, the Hubbard model, alongside applications to molecular systems.
The Hamiltonians are built using Qiskit~\cite{javadi2024quantum}, and
all results are generated using the Hamming-weight sampling strategy. For all systems, with random seeds ranging from 0 to 99, we use a subsystem solver with a width of 8 (or the system size, if smaller) and a depth of 4. For the 1D and 2D hopping models and the Hubbard model, we report the results obtained using either the Jordan–Wigner or Bravyi–Kitaev transformation, with a maximum of 30,000 iterations. Figure~\ref{fig:1d-20-hopping} to Figure~\ref{fig:hubbard2d} are reporting the best result across all runs to match the set up in the previous paper~\cite{yu2025clifford}, and the mean $\pm$ standard deviation results are shown in Figure~\ref{fig:latticePMstd}. For the molecular systems, we consistently use the Jordan–Wigner transformation and limit the number of iterations to 2,000. Table~\ref{tab:mole-sto-3g} and Table~\ref{tab:mole-6-31g} show the best results under the STO-3G and 6-31G basis, while the best percentage reduction is shown in Figure~\ref{fig:mole}. More detailed results and their corresponding runtime can be found in the Supporting Information.

We report the total Pauli weights for all the models tested using RSD, and the weighted Pauli weights for the molecular systems. For an operator $O \in \mathcal{H}_f$, we show the improvement of the mapper generated by RSD, which is called $\Phi_{\text{RSD}}$, over a given mapping $\Phi$ by the percentage reduction defined as
\begin{equation}
    \text{PR}_{\Phi} := 1 - \frac{\text{PW}(\Phi_{\text{RSD}}(O))}{\text{PW}(\Phi(O))}.
\end{equation}
Similarly, $\text{PR}^{\text{w}}_{\Phi}$ is used for computing weighted Pauli weight reductions.
We will use $\Phi_{\text{Conv}}$ as the best among the conventional mappers, say the Jordan-Wigner, Bravyi-Kitaev and ternary tree mapping. As for the mapping produced by simulated annealing, we would call that as $\Phi_{\text{Annealing}}$.

\subsection{Lattice hopping models}
\label{sec:lattice}

Lattice models~\cite{fradkin2013field, lieb1961two} serve as fundamental paradigms for studying strongly correlated materials~\cite{leblanc2015solutions}, quantum magnetism~\cite{spinelli2015atomically}, and programmable quantum emulators~\cite{altman2021quantum}. Numerical simulations of these models are of central importance, providing powerful means to reveal novel physical phenomena and exotic phases of matter. In this section, we present numerical experiments on hopping models to benchmark the performance of the RSD algorithm, with particular emphasis on solution quality and scalability.

\subsubsection{1D lattice hopping}
\begin{figure*}[ht!]
    \centering
    \includegraphics[width=0.49\linewidth]{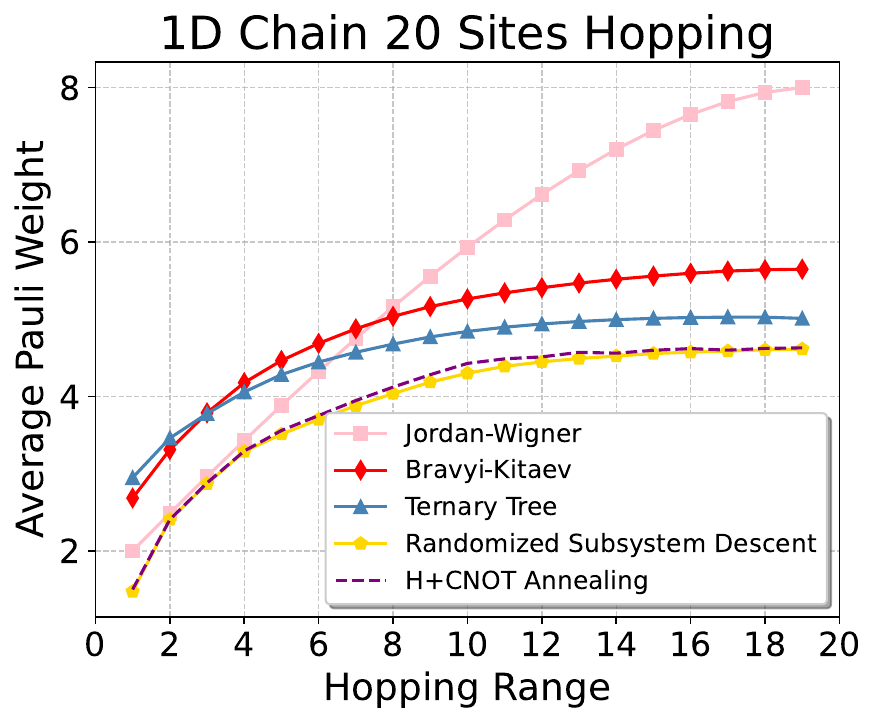}
    \includegraphics[width=0.49\linewidth]{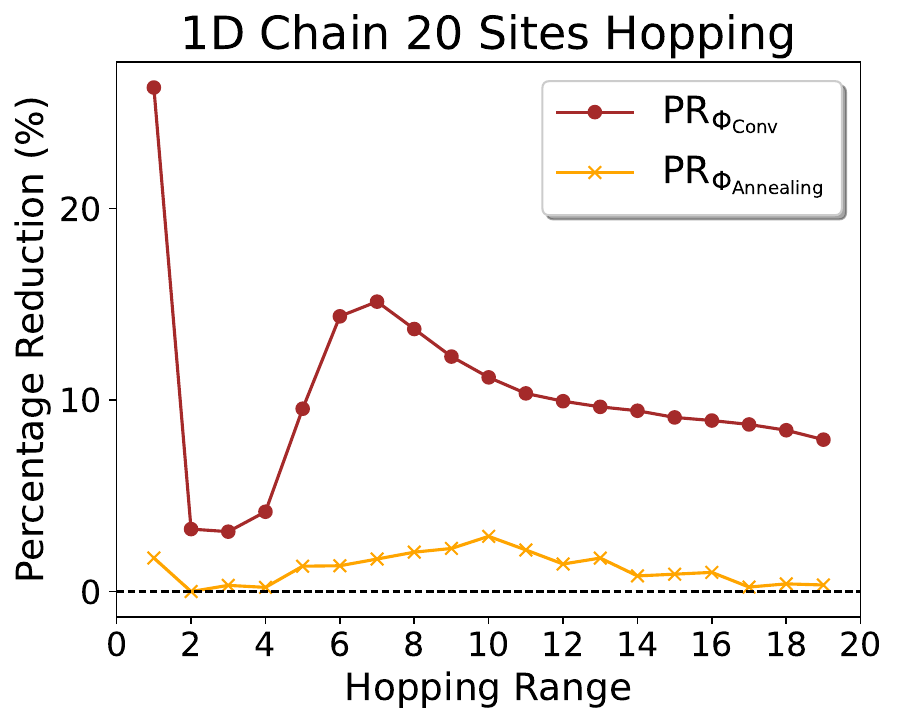}
    \caption{1D lattice hopping with varying interaction range as shown in Equation~\ref{eqn:1d-hopping}. Here $N = 20$ and $r$ ranges from 1 to 19. Left: average Pauli weight of Jordan-Wigner, Bravyi-Kitaev, ternary tree, H+CNOT annealing and randomized subsystem descent. Right: percentage reduction of $\text{PR}_{\Phi_{\text{Conv}}}$ and $\text{PR}_{\Phi_{\text{Annealing}}}$. The results achieved by RSD perform at least equally well compared with H+CNOT annealing.}
    \label{fig:1d-20-hopping}
\end{figure*}

To rigorously benchmark our method, we first consider the hopping model on a one-dimensional lattice, defined by the Hamiltonian
\begin{equation}\label{eqn:1d-hopping}
    H = \sum_{0 < |i - j| \leq r}^N a_i^\dag a_j.
\end{equation}
We aim to directly compare our performance against the recent Clifford circuit optimization framework~\cite{yu2025clifford} that used simulated annealing, which reports the best result among multiple runs. For a 20-site system, we varied the hopping range $r$ from 1 to 19, with the best result among all random seeds detailed in Figure~\ref{fig:1d-20-hopping}.

\begin{figure*}[htbp]
    \centering
    \includegraphics[width=0.49\linewidth]{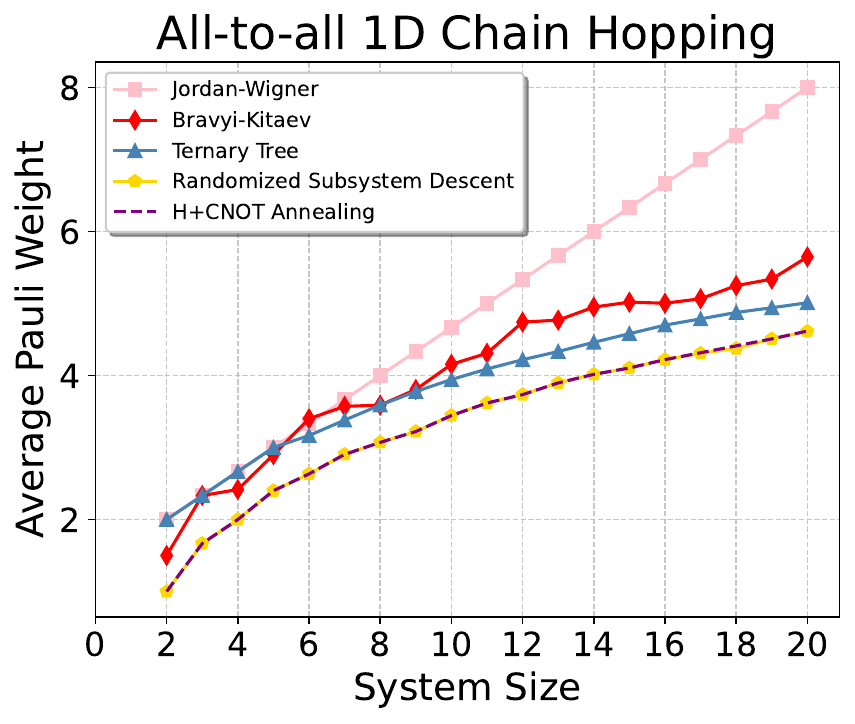}
    \includegraphics[width=0.49\linewidth]{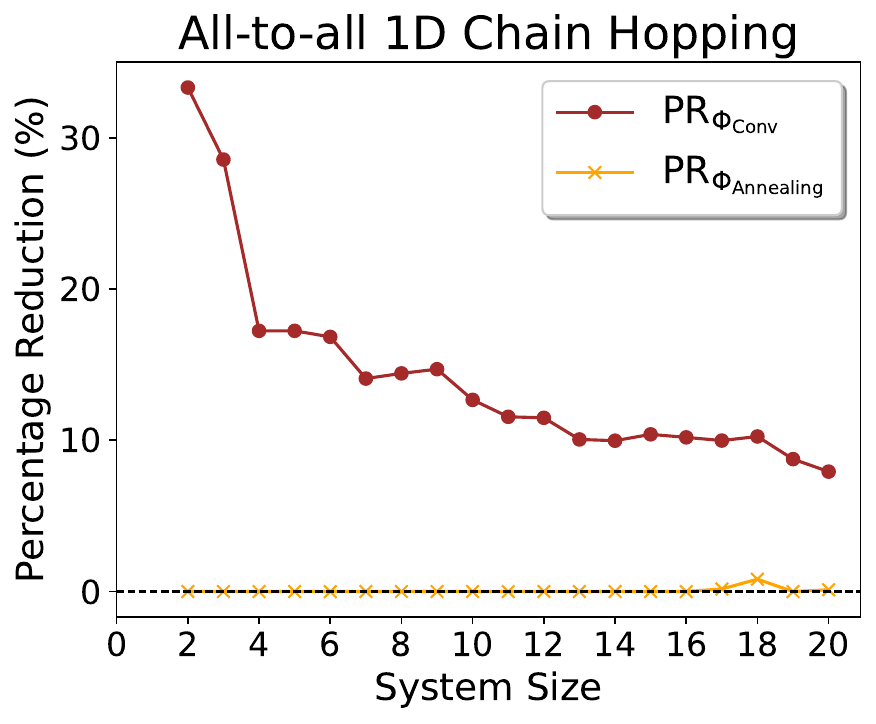}
    \caption{All-to-all 1D lattice hopping with varying system size as shown in Equation~\ref{eqn:1d-hopping-all-to-all}. Here $N$ ranges from 2 to 20. Left: average Pauli weight of Jordan-Wigner, Bravyi-Kitaev, ternary tree, H+CNOT annealing and randomized subsystem descent. Right: percentage reduction of $\text{PR}_{\Phi_{\text{Conv}}}$ and $\text{PR}_{\Phi_{\text{Annealing}}}$. The results achieved by RSD perform at least equally well compared with H+CNOT annealing.}
    \label{fig:1d-all-to-all}
\end{figure*}

Our optimized Hamiltonians universally outperform those generated by the conventional mappers. The magnitude of this advantage peaks at intermediate hopping ranges ($r \in \{6,7,8\}$) before gradually tapering to a 10\% reduction as the interaction range increases.

When compared against the resource-intensive simulated annealing approach~\cite{yu2025clifford}, RSD yields superior optimization outcomes across more than half of the evaluated data points, where at most 1.5\% improvement is achieved. For the specific case of nearest-neighbor hopping ($r=1$), our algorithm efficiently and precisely recovers the optimal Hamiltonian~\cite{fradkin2013field}
\begin{equation}
\begin{split}
    H = X_1 &+ \sum_{i=1}^{9} (Z_i Z_{i+1} + X_{i+1}) + X_{11} \\& + \sum_{i=11}^{19} (Z_i Z_{i+1} + X_{i+1}).
\end{split}
\end{equation}
This exact structural recovery drops the average Pauli weight from $2.0$ (Jordan-Wigner) to $1.47$, perfectly matching the global minimum. However, for the simulated annealing method~\cite{yu2025clifford}, it only found the solution that leads to a $25\%$ drop (from 2.0 to 1.5), making the final result sub-optimal.

We further challenge our algorithm with the 1D all-to-all hopping model, where the Hamiltonian is written as
\begin{equation}\label{eqn:1d-hopping-all-to-all}
    H = \sum_{0 < |i - j| \leq {N-1}}^N a_i^\dag a_j.
\end{equation}
Because this system implies full connectivity, it exhibits fewer local structures that can be easily exploited by fermion-to-qubit mappings.

The comparative results, shown in Figure~\ref{fig:1d-all-to-all}, explicitly highlight the quality of the solution found by our algorithm. It can be easily seen that the RSD method is better than all conventional mappers and performs at least equally well as the simulated annealing. In addition, RSD achieves higher Pauli-weight reductions than simulated annealing for system sizes larger than $N=15$. 

These results demonstrate that, despite lacking a theoretical guarantee of global optimality, our greedy subsystem approach is able to provide high quality results under limited computational budgets.

\subsubsection{2D lattice hopping}

\begin{figure*}[t!]
    \centering
    \includegraphics[width=0.49\linewidth]{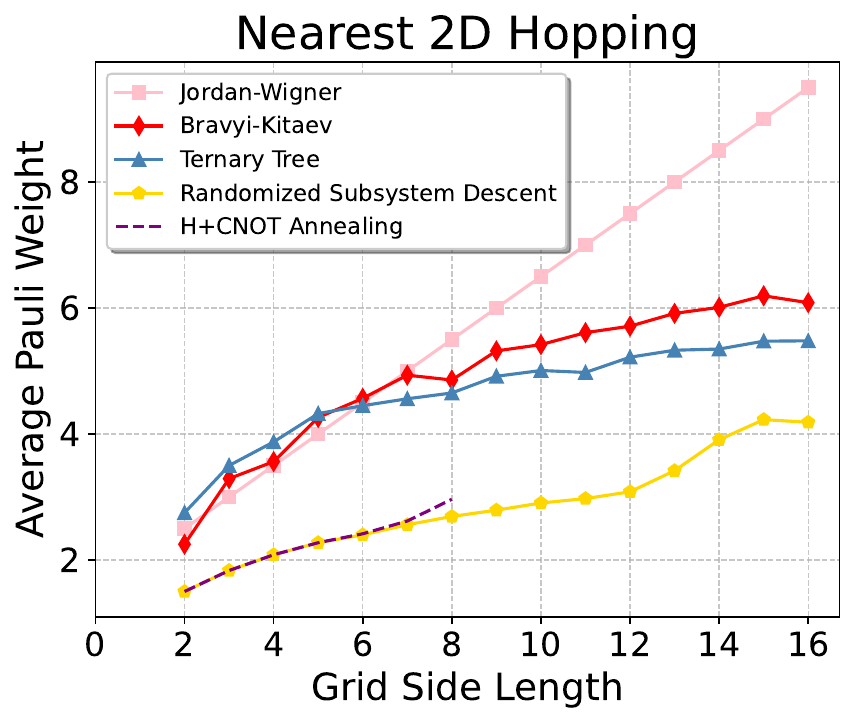}
    \includegraphics[width=0.49\linewidth]{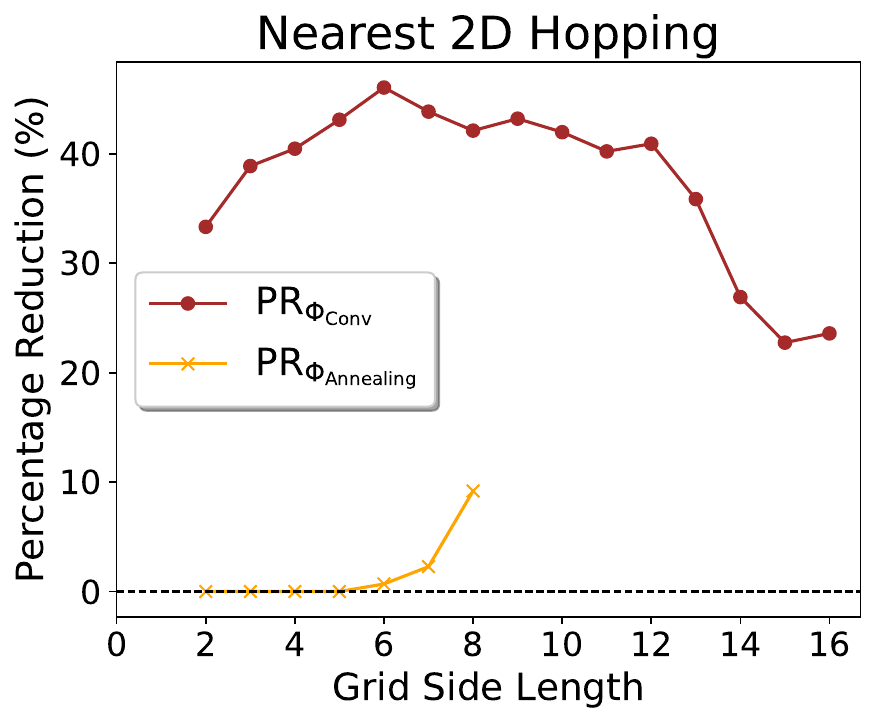}
    \caption{Nearest neighbor 2D lattice hopping with varying system size as shown in Equation~\ref{eqn:2d-nearest-neighbor}. Here $N$ ranges from 2 to 16. Left: average Pauli weight of Jordan-Wigner, Bravyi-Kitaev, ternary tree, H+CNOT annealing ($N$ ranges from $2$ to $8$) and randomized subsystem descent. Right: percentage reduction of $\text{PR}_{\Phi_{\text{Conv}}}$ and $\text{PR}_{\Phi_{\text{Annealing}}}$. The results achieved by RSD perform equally well compared with H+CNOT annealing for $N = 2,3,4,5,6$, and achieves noticeable improvements for $N = 7$ and $8$.}
    \label{fig:2d-nearest-hopping}
\end{figure*}

Now we turn to investigate 2D hopping models.
Because they introduce non-trivial spatial connectivity compared to one-dimensional chains, optimizing the fermion-to-qubit mapping for these systems is significantly more challenging.

For the nearest-neighbor hopping model on a 2D lattice, the Hamiltonian is given by
\begin{equation}
\label{eqn:2d-nearest-neighbor}
    H = \sum_{\langle i,j \rangle} a_i^\dag a_j
\end{equation}
where $\langle i,j \rangle$ denotes all pairs of nearest-neighbor sites. The optimization results are presented in Figure~\ref{fig:2d-nearest-hopping}, evaluating grid side lengths ranging from 2 to 16. For a 2D system of grid side length $n$, the system size scales as $\mathcal{O}(n^2)$. This rapid scaling renders global optimization techniques, such as simulated annealing, computationally intractable due to the quickly expanding action space. In contrast, the RSD method exhibits excellent scalability. Compared with conventional mappers, our method consistently achieves a Pauli-weight reduction of over 30\% for grid side lengths up to 12, and approximately a 20\% improvement for larger grids. 

Notably, for the 2D nearest-neighbor hopping model, employing the Bravyi-Kitaev transformation as the initial mapping yields superior optimization outcomes as the grid side length increases. This behavior may be due to the property that the Bravyi-Kitaev mapping captures higher-dimensional spatial locality more effectively than standard Jordan-Wigner mappings~\cite{o2024ultrafast}.

\subsection{Hubbard model}
\label{sec:hubbard}

\begin{figure*}[htbp]
    \centering
    \includegraphics[width=0.49\linewidth]{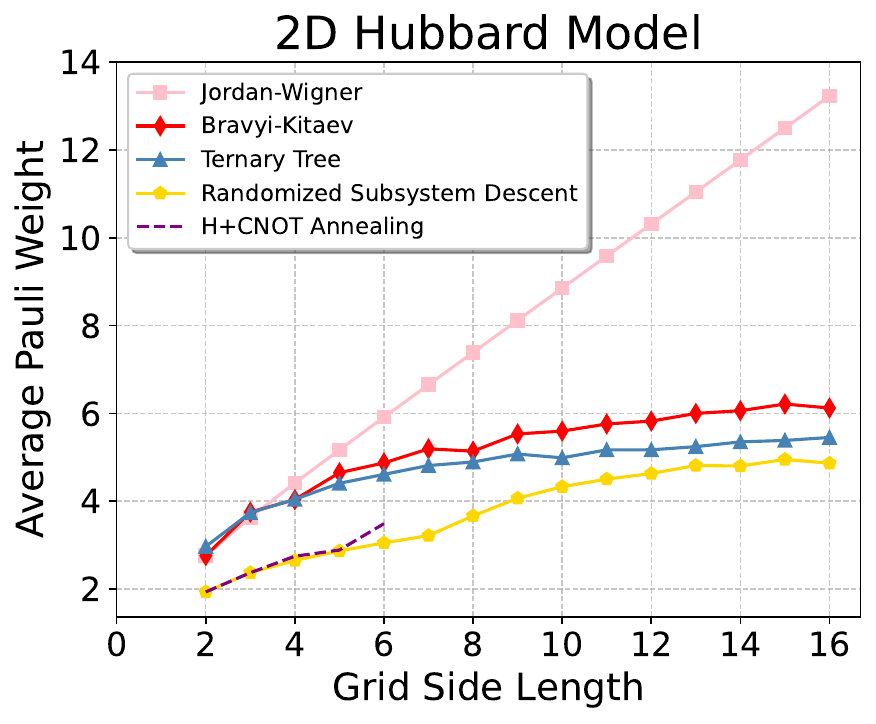}
    \includegraphics[width=0.49\linewidth]{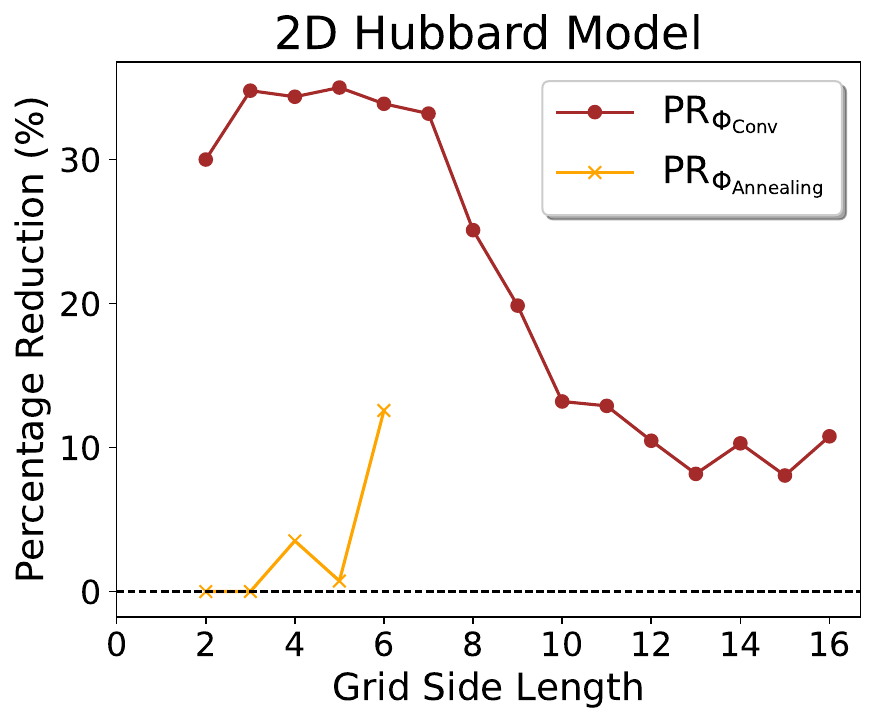}
    \caption{2D Hubbard model with varying system size as shown in Equation~\ref{eqn:hubbard}. Here grid side length $N$ ranges from 2 to 16. Left: average Pauli weight of Jordan-Wigner, Bravyi-Kitaev, ternary tree, H+CNOT annealing ($N$ ranges from $2$ to $6$) and randomized subsystem descent. Right: percentage reduction of $\text{PR}_{\Phi_{\text{Conv}}}$ and $\text{PR}_{\Phi_{\text{Annealing}}}$. The results achieved by RSD perform equally well compared with H+CNOT annealing for $N = 2,3$, and achieves more than $10\%$ improvement for $N = 6$.}
    \label{fig:hubbard2d}
\end{figure*}

The Hubbard model is foundational in condensed matter physics, serving as essential paradigms for investigating strongly correlated phenomena such as the quantum Hall effect and high-temperature superconductivity~\cite{barends2015digital, leblanc2015solutions}.
Here the Hamiltonian is defined as
\begin{equation}
\label{eqn:hubbard}
H = -t\sum_{\langle i, j \rangle, \sigma} \left( a_{i,\sigma}^\dagger a_{j,\sigma} + a_{j,\sigma}^\dagger a_{i,\sigma} \right) + U \sum_{i} n_{i,\uparrow} n_{i,\downarrow},
\end{equation}
where $\langle i,j \rangle$ denotes all pairs of nearest-neighbor sites, $\sigma$ is the spin, and $n_{i,\sigma} = a_{i,\sigma}^\dag a_{i,\sigma}$.

A comparison with conventional mappers and the simulated annealing method is shown in Figure~\ref{fig:hubbard2d}.
For grid side lengths up to 6, the RSD algorithm achieves a substantial Pauli-weight reduction of over 30\% relative to conventional mappers, and outpaces simulated annealing by more than 10\%. Furthermore, our methodology readily scales to larger systems: for grid side lengths exceeding 10, the algorithm sustains its computational advantage, yielding improvements of up to 10\% over the highly optimized ternary tree mapping at the $16\times16$ case.

\subsection{Electronic structure models}
\label{sec:mols}
The accurate simulation of electronic structures represents a cornerstone of computational chemistry and a primary target for near-term quantum algorithms~\cite{szabo2012modern,lanyon2010towards,babbush2014adiabatic}. Consequently, optimizing the fermion-to-qubit mapping for these highly intricate systems is essential for minimizing quantum resource overhead.
The electronic structure Hamiltonian can be modeled as
\begin{equation}
    H = \sum_{i,j} h_{ij}a_i^\dag a_j + \sum_{i,j,k,\ell} h_{ijk\ell} a_i^\dag a_j^\dag a_\ell a_{k},
\end{equation}
where $h_{ij}$ and $h_{ijk\ell}$ are determined by the one- and two-electron integrals, respectively. All fermionic Hamiltonians utilized in our benchmarks were generated using the PySCF package~\cite{sun2018pyscf,sun2020recent,sun2026python}. 
Unlike the lattice hopping models discussed previously, molecular Hamiltonians are more complicated because of the two-electron interaction terms. Furthermore, these Hamiltonians would be mapped into a substantially larger number of distinct Pauli strings. This makes the optimization over weighted Pauli weight more reasonable, since algorithms like qDRIFT~\cite{campbell2018random} would lead to a more efficient Hamiltonian simulation implementation.

We benchmarked the performance of our RSD algorithm using the metrics of total Pauli weight and weighted Pauli weight, against conventional mapping strategies and the recently developed Hamiltonian-Adaptive Ternary Tree (HATT) mapper~\cite{liu2025hatt}, which attains the specific structure of the target Hamiltonian during the mapping process.

  \begin{table*}[htbp]
    \centering
    \caption{(Weighted) Pauli weight for different molecules in the
    STO-3G basis. Here modes represent the number of fermionic modes,
    \#PS is the number of Pauli strings. JW represents the Jordan-Wigner
    transformation, BK represents the Bravyi-Kitaev transformation,
    TT represents the ternary tree mapping, HATT represents the result
    from Hamiltonian-Adaptive ternary tree mapping~\cite{liu2025hatt},
    and RSD represents randomized subsystem descent.
    PW stands for the total Pauli weight and wPW represents the
    coefficient-weighted Pauli weight.
    RSD entries are the best outcomes out of 
    100 seeds (0--99).}
    \label{tab:mole-sto-3g}
    \resizebox{\textwidth}{!}{
    \begin{tabular}{ccc|cc|cc|cc|cc|cc}
    \hline
    System & modes & \textbf{\#}PS
    & \multicolumn{2}{c|}{JW}
    & \multicolumn{2}{c|}{BK}
    & \multicolumn{2}{c|}{TT}
    & \multicolumn{2}{c|}{HATT}
    & \multicolumn{2}{c}{RSD} \\
    & & & PW & wPW & PW & wPW & PW & wPW & PW & wPW & PW & wPW \\
    \hline
$\mathrm{H}_2$ & 4 & 15
& 32 & 3.355
& 34 & 4.403
& 36 & 4.156
& 32 & 3.355
& \textbf{26} & \textbf{3.192} \\
$\mathrm{LiH}$ & 12 & 631
& 3248 & 28.882
& 3660 & 39.561
& 3536 & 46.820
& 2850 & 33.924
& \textbf{2752} & \textbf{25.969} \\
$\mathrm{H}_2\mathrm{O}$ & 14 & 1086
& 6332 & 158.681
& 6567 & 235.926
& 6658 & 197.121
& 5545 & 194.899
& \textbf{5139} & \textbf{134.355} \\
$\mathrm{CH}_4$ & 18 & 5956
& 42476 & 214.488
& 42646 & 268.266
& 41530 & 284.462
& 38733 & 239.914
& \textbf{34719} & \textbf{179.359} \\
$\mathrm{O}_2$ & 20 & 2239
& 16904 & 345.992
& 16828 & 473.707
& 15364 & 451.018
& 13082 & 382.250
& \textbf{12108} & \textbf{298.907} \\
$\mathrm{NaF}$ & 28 & 24275
& 247264 & 1224.228
& 217827 & 1416.652
& 206710 & 1419.324
& 192072 & 1346.403
& \textbf{168786} & \textbf{927.003} \\
$\mathrm{CO}_2$ & 30 & 16170
& 173324 & 919.296
& 144112 & 1091.087
& 138756 & 1074.355
& 131992 & 1077.890
& \textbf{114934} & \textbf{704.021} \\
\hline
    \end{tabular}
    }
  \end{table*}

  \begin{table*}[htbp]
    \centering
    \caption{The same as Table~\ref{tab:mole-sto-3g}, but in the 6-31G basis.}
    \label{tab:mole-6-31g}
    \resizebox{\textwidth}{!}{
    \begin{tabular}{ccc|cc|cc|cc|cc|cc}
    \hline
    System & modes & \textbf{\#}PS
    & \multicolumn{2}{c|}{JW}
    & \multicolumn{2}{c|}{BK}
    & \multicolumn{2}{c|}{TT}
    & \multicolumn{2}{c|}{HATT}
    & \multicolumn{2}{c}{RSD} \\
    & & & PW & wPW & PW & wPW & PW & wPW & PW & wPW & PW & wPW \\
    \hline
$\mathrm{H}_2$ & 8 & 185
& 728 & 26.009
& 756 & 36.313
& 834 & 33.199
& 768 & 30.588
& \textbf{664} & \textbf{23.882} \\
$\mathrm{LiH}$ & 22 & 8758
& 73872 & 159.928
& 72353 & 194.250
& 65586 & 177.067
& 61576 & 169.181
& \textbf{54525} & \textbf{124.285} \\
$\mathrm{H}_2\mathrm{O}$ & 26 & 12732
& 123060 & 828.321
& 110562 & 856.438
& 104934 & 834.224
& 88244 & 751.954
& \textbf{82378} & \textbf{565.152} \\
$\mathrm{CH}_4$ & 34 & 75444
& 899480 & 1370.501
& 715392 & 1305.590
& 667968 & 1200.171
& 621970 & 1168.109
& \textbf{585977} & \textbf{902.592} \\
$\mathrm{O}_2$ & 36 & 22543
& 284280 & 1906.171
& 224972 & 1849.149
& 207120 & 1773.386
& 185398 & 1718.265
& \textbf{166706} & \textbf{1253.402} \\
$\mathrm{NaF}$ & 44 & 157607
& 2347244 & 3257.456
& 1791496 & 2981.287
& 1587055 & 2742.956
& 1466451 & 2672.994
& \textbf{1376849} & \textbf{1907.039} \\
$\mathrm{CO}_2$ & 54 & 182270
& 3234356 & 6636.245
& 2172261 & 5191.622
& 1950795 & 4829.675
& 1836844 & 4626.048
& \textbf{1741234} & \textbf{3739.292} \\
\hline
    \end{tabular}
    }
  \end{table*}

\begin{figure*}[htbp]
    \centering
    \includegraphics[width=0.75\linewidth]{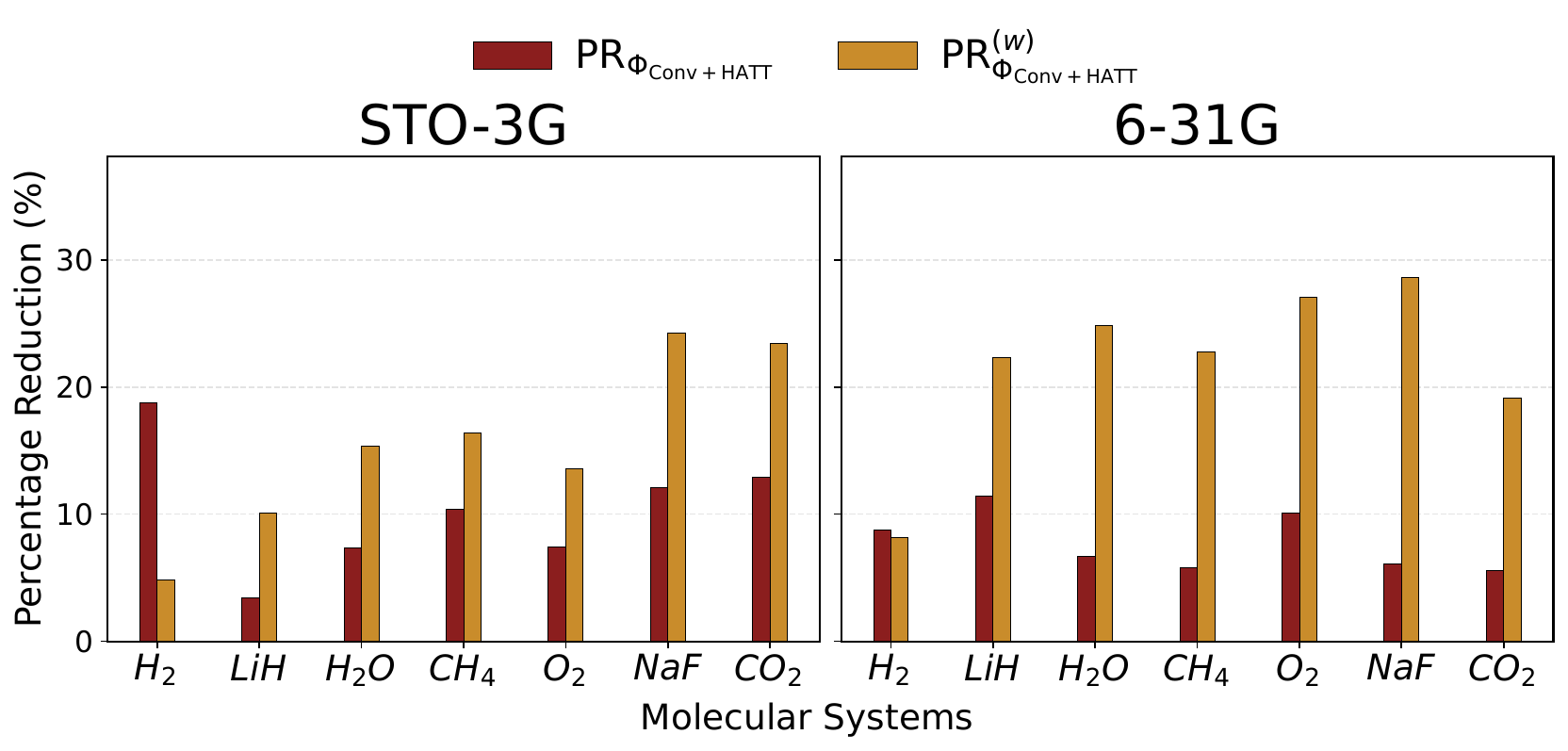}
    \caption{Best percentage reduction across all random seeds achieved by the RSD algorithm for the molecular systems. $\text{PR}_{\Phi_{\text{Conv+HATT}}}$ is the improvement of the total Pauli weight, and $\text{PR}^{(\text{w})}_{\Phi_{\text{Conv+HATT}}}$ is the improvement on weighted Pauli weight, compared with the best out of Jordan-Wigner, Bravyi-Kitaev, balanced ternary tree and the HATT mapper. }
    \label{fig:mole}
\end{figure*}

Across all the molecular systems, the optimization procedure was capped for at most 2000 total iterations. It should be emphasized that this configuration represents a very modest allocation of computational resources. Our comparative results are summarized in Table~\ref{tab:mole-sto-3g} and Table~\ref{tab:mole-6-31g}, with the corresponding percentage reductions in both Pauli weight and weighted Pauli weight visualized in Figure~\ref{fig:mole}. The results showing the mean and sample deviation can be seen in the Supporting Information.
It should be noted that the increase in the number of Pauli strings introduces considerable computational overhead when updating the Pauli strings during each optimization step, thereby lengthening the runtime required to execute a subsystem update.

Although optimization reduces the total or weighted Pauli weight of the Hamiltonian, it may increase the Pauli weights of symmetry observables, such as particle number and the $z$-component of spin, as shown in the Supporting Information. This tradeoff arises because their Pauli weights are not included in the optimization objective. The benefits of reducing the Hamiltonian Pauli weight should therefore be assessed alongside the cost of measuring these symmetry observables.

Despite these computational challenges, the RSD algorithm demonstrated robust performance. For systems evaluated in both basis sets (STO-3G and 6-31G), our method universally achieved a total Pauli weight reduction relative to the compared mappers, even for the highly specialized HATT mapper. For the weighted Pauli weight, RSD produces substantial reductions, especially in the 6-31G basis. This is due to the fact that previous fermion-to-qubit mapping designs do not allow problem-aware optimization, thus their performance will not be as good.

\section{Conclusions and outlook}
\label{sec:conclusions}
In this work, we introduce Randomized Subsystem Descent as an efficient algorithm framework for optimizing Hamiltonian encodings. Inspired by the principles of randomized block coordinate descent, RSD employs a subsystem solver to iteratively minimize the global objective function. By confining the optimization to a small subset of qubits at each step, the greedy acceptance criterion guarantees monotonic improvement. The algorithm demands modest computational resources, and the numerical experiments presented herein were executed within hours without the need for highly specialized software engineering. Maintaining this low classical preprocessing overhead ensures that the theoretical quantum speedup in downstream Hamiltonian simulations is preserved.

Several promising directions for future work remain to be explored. First, since the (weighted) Pauli weight can only be seen as an indicator of potential hardware-efficient quantum algorithms, one may leverage
the objective-agnostic architecture of the algorithmic framework to seamlessly substitute the current objectives with alternative cost functions~\cite{liu2025hatt,steudtner2019quantum,leone2024practical, d2023challenges}. Optimizing for the maximum Pauli weight, locality, circuit depth, or hardware-aware metrics governed by specific qubit connectivity topologies could yield customized mappings tailored for targeted quantum architectures. 

Second, while our brute-force Clifford solver is highly efficient for small subsystems, the exponential scaling of the search space inherently bounds the reach of each iteration. Scaling to larger local dimensions will necessitate the development of more sophisticated, structure-aware subsystem solvers or the integration of heuristic search-tree pruning to maintain computational tractability. Moreover, the current optimization is strictly confined to the Clifford group. Expanding the allowed operation space to include non-Clifford transformations, or combining the method with continuous optimization paradigms~\cite{rudolph2025pauli}, could unlock previously inaccessible mappings.

Third, refining the subsystem sampling process presents a critical opportunity for algorithmic acceleration. While our current sampling strategies demonstrate strong empirical success, deploying advanced machine learning techniques, particularly reinforcement learning agents or Markov chain-based state-aware policies could drastically enhance convergence rates by dynamically learning to target high-impact subsystems. 

Finally, establishing a rigorous theoretical foundation for randomized subsystem descent remains an open challenge. Translating the well-established convergence guarantees of continuous block coordinate descent~\cite{nesterov2012efficiency,xu2013block,xu2017globally} to the discrete space of Pauli-weight minimization is nontrivial. Formally establishing convergence rates, even under simplifying structural assumptions, would place our empirical findings on a firmer theoretical footing and actively guide the design of future sampling distributions.

In conclusion, Randomized Subsystem Descent provides a highly scalable, practical, and mathematically extensible framework for finding efficient fermion-to-qubit mappings. By strictly bounding the per-iteration computational cost while ensuring monotonic descent, its simplicity and robust empirical performance make it a candidate algorithmic framework for suppressing simulation overhead for quantum hardware.

\section{Supporting Information}
Please see the Supporting Information for
(i) Pauli weights for particle number and the $z$-component of spin before and after the optimization for the molecular systems;
(ii) Wall-clock timings, final objective values, and their visualizations, with RSD for all systems tested across this manuscript;
(iii) Comparisons of costs, final objective values of different sampling methods as discussed in Section~\ref{sec:rsd} and Appendix~\ref{append:sampling}.

\section{Data and Software Availability}
The data underlying this study are available in the public GitHub repository at 
\href{https://github.com/Genz17/F2Q}{https://github.com/Genz17/F2Q}.
The scripts and code used to generate and analyze the results are available in the same repository.

\section{Acknowledgement}
We thank Tianlong Chen, Rana Muhammad Shahroz Khan, Pingzhi Li and Krishnan Raghavan for insightful discussions. We are also grateful to Jeffery Yu for providing the simulated annealing data, and to both him and Hong-Zhou Ye for their constructive feedback on the manuscript.
G.Y. and X.W. were supported by the U.S. Department of Energy, Office of Science, Accelerated Research in Quantum Computing Centers, Quantum Utility through Advanced Computational Quantum Algorithms, grant no. DE-SC0025341, and the U.S. National Science Foundation grant CCF-1942837 (CAREER). X.W. was also supported by a Sloan research fellowship. 
H.Y. was partially supported by the US National Science Foundation under awards IIS-2520978, GEO/RISE5239902, DOE (ASCR) Award DE-SC0026052, and
the DARPA D24AP00325-00.
J.L. was supported by the DOE-SC Office of Advanced Scientific Computing Research MACH-Q project under contract number DE-AC02-06CH11357. 

\bibliography{ref}


\appendix

\section{Clifford transformations}
\label{append:clifford}

One important type of unitary operators is the set of Pauli operators, namely
\begin{equation}
\begin{split}
    I = 
    \begin{bmatrix}
        1 & \\
         & 1
    \end{bmatrix}, &\;
    X = 
    \begin{bmatrix}
     & 1\\
    1 & 
    \end{bmatrix},\\
    Y = 
    \begin{bmatrix}
     & -\imath \\
    \imath & 
    \end{bmatrix}, &\;
    Z = 
    \begin{bmatrix}
    1 &  \\
     & -1
    \end{bmatrix}.
\end{split}
\end{equation}
They form a basis of all the linear operators acting on $\mathbb{C}^2$.
Conventionally we would write $X_i$ for a multi-qubit Hamiltonian to indicate the
tensor product of multiple identity matrices while the $i$-th operand is $X$. The same notation is also adopted for $Y_i$ and $Z_i$.

Consider the Clifford gates $\{\text{CNOT}, H, S\}$, where
\begin{equation}
\begin{split}
    \text{CNOT} = 
    \begin{bmatrix}
        1 \\
        & 1\\
        & & & 1\\
        & & 1 & 
    \end{bmatrix},\\
    H = \frac{1}{\sqrt{2}}\begin{bmatrix}
        1 & 1\\
        1& -1
    \end{bmatrix},
    S = \begin{bmatrix}
        1 & \\
        & \imath
    \end{bmatrix},
\end{split}
\end{equation}
we know for the single qubit gates $H$ and $S$,
\begin{equation}
\begin{split}
    H^\dag X H = Z, \quad
    H^\dag Y H &= -Y, \quad
    H^\dag Z H = X,\\
    S^\dag X S = -Y, \quad
    S^\dag Y S &= X, \quad
    S^\dag Z S = Z.
\end{split}
\end{equation}
As for the CNOT gate, a complete unitary conjugation table for the Paulis are presented in Table~\ref{tab:cnot_pauli_conj}.
\begin{table}[ht!]
\centering
\caption{Conjugation of all two-qubit Pauli strings under the CNOT gate, where the signs are dropped. Entries highlighted in red indicate weight reductions.}
\label{tab:cnot_pauli_conj}
\setlength{\tabcolsep}{5pt}
\renewcommand{\arraystretch}{1.2}
\begin{adjustbox}{max width=\columnwidth}
\begin{tabular}{c c c c c c c c c}
\toprule
\textbf{Input} 
& \ptwo{I}{I} & \ptwo{I}{X} & \ptwo{I}{Y} & \ptwo{I}{Z}
& \ptwo{X}{I} & \hlptwo{X}{X} & \ptwo{X}{Y} & \ptwo{X}{Z} \\
\textbf{Output} 
& \ptwo{I}{I} & \ptwo{I}{X} & \ptwo{Z}{Y} & \ptwo{Z}{Z}
& \ptwo{X}{X} & \hlptwo{X}{I} & \ptwo{Y}{Z} & \ptwo{Y}{Y} \\
\midrule
\textbf{Input} 
& \ptwo{Y}{I} & \hlptwo{Y}{X} & \ptwo{Y}{Y} & \ptwo{Y}{Z}
& \ptwo{Z}{I} & \ptwo{Z}{X} & \hlptwo{Z}{Y} & \hlptwo{Z}{Z} \\
\textbf{Output} 
& \ptwo{Y}{X} & \hlptwo{Y}{I} & \ptwo{X}{Z} & \ptwo{X}{Y}
& \ptwo{Z}{I} & \ptwo{Z}{X} & \hlptwo{I}{Y} & \hlptwo{I}{Z} \\
\bottomrule
\end{tabular}
\end{adjustbox}
\end{table}

\begin{figure*}[htbp]
    \centering
    \includegraphics[width=1.0\linewidth]{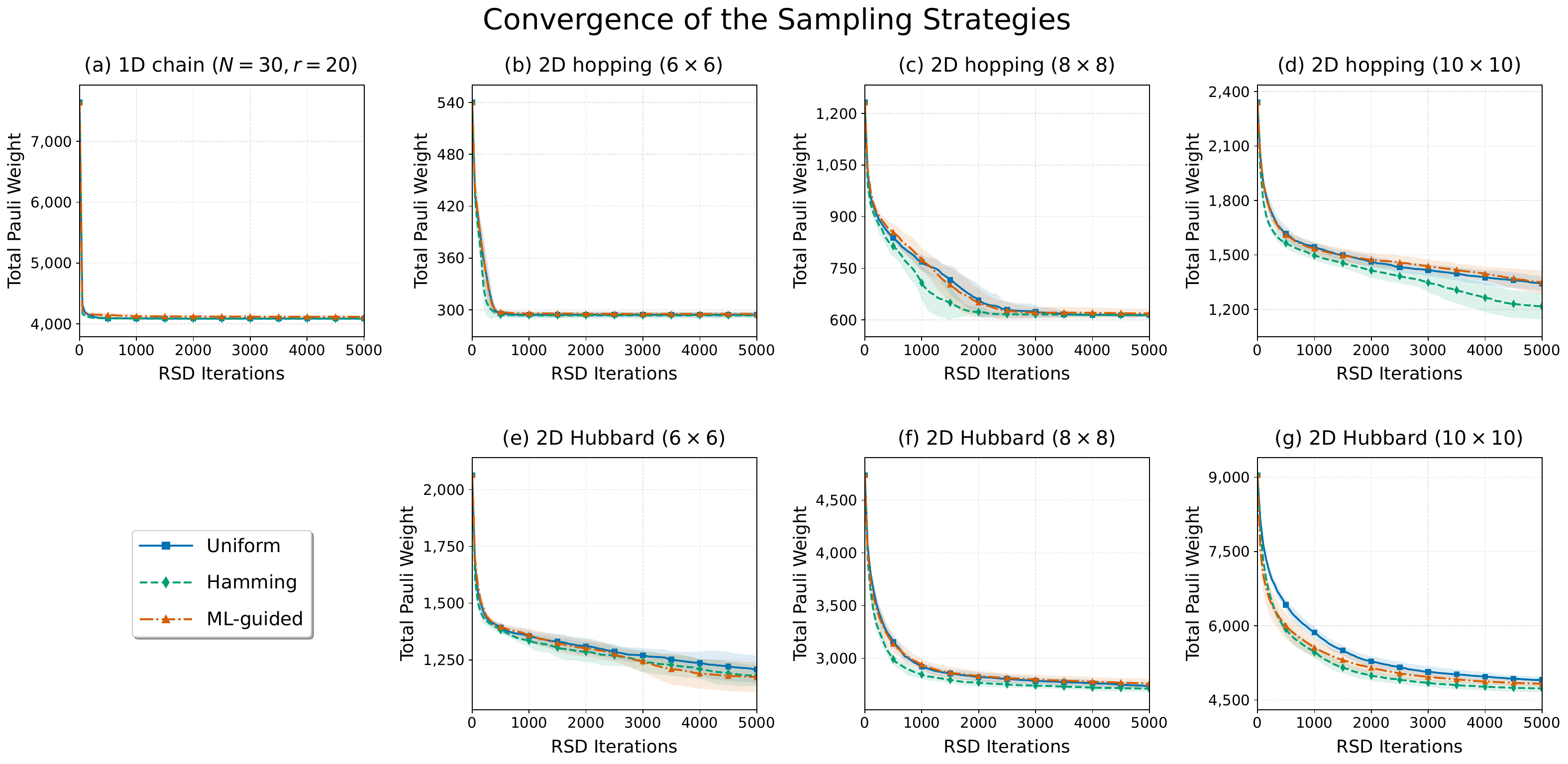}
    \caption{The convergence behavior of the three different sampling methods on the tested systems.}
    \label{fig:3-sampling}
\end{figure*}

\section{Comparison of sampling methods}
\label{append:sampling}
In this section we conduct numerical investigations to different sampling methods.
The three different sampling methods mentioned in the main text are the uniform sampling strategy, Hamming weight sampling strategy and the ML-induced sampling strategy. We assume the global system with $N_M$ Pauli strings is of size $n$ and the subsystem is of size $k$.
For the uniform sampling, one chooses $k$ qubits out of the $n$ choices with equal possibilities. For the Hamming weight sampling strategy, one needs to first compute the Hamming weight $h_i$ at qubit index $i$, defined as the number of Pauli strings that act nontrivially on qubit 
$i$, that is, the number of non-identity operators appearing at that qubit across all Pauli strings. Since $h_i \geq 0$, we can sample $k$ qubits with the probability distribution 
\begin{equation}
    P_i = \frac{h_i+\epsilon}{\left(\sum_{j}h_j\right) + n\epsilon},
\end{equation}
where $\epsilon$ is a small constant for avoiding zero probability at some index. For the machine-learning-induced sampling strategy, given a fixed subsystem solver, one can construct a dataset in which the input consists of Hamiltonian features and the label corresponds to the Pauli-weight reduction achieved by the subsystem solver. Even with a lightweight supervised learning model trained on a relatively small dataset, the model is able to capture informative underlying patterns that correlate with potential Pauli-weight reductions. When applying for inference, one may firstly sample multiple different choices of $k$ qubits out of the giant $\binom{n}{k}$ spaces, use the trained model to predict the expected reduction for each candidate, and finally apply the subsystem solver to the subset predicted to yield the largest reduction.

We investigated ML-guided subsystem sampling using five XGBoost
  regressors trained with different random seeds. Each model contained
  300 trees with maximum depth 8 and learning rate 0.05, and was trained
  on 30-site one-dimensional hopping chains with
  $r\in\{10,12,14,16,18,22,24,26,28\}$.
  For each chain, ten restarts of 150 uniformly sampled RSD steps
  provided 13,500 training examples per model, labeled by the achieved
  local Pauli-weight reduction. During inference, the model ranked
  pools of 200 randomly sampled candidate subsystems by their predicted
  reduction. Without retraining, we compared this strategy with uniform
  and Hamming-based sampling on the held-out chain with $N=30$, $r=20$,
  and on two-dimensional nearest-neighbor hopping and Hubbard models
  on $6\times6$, $8\times8$, and $10\times10$ grids.
  All optimizations started from Jordan--Wigner and used subsystem
  width $k=8$, local depth $d=4$ for 5000 iterations.
  Figure~\ref{fig:3-sampling} reports the optimization behavior
  over ten optimization seeds for each baseline.
  The benefit of ML guidance depended on the system and optimization
  stage. In the $10\times10$ Hubbard model, the mean PW reduction after
  100 iterations was $22.3\%$ for ML-guided sampling, compared with
  $18.9\%$ for Hamming-based sampling and $15.4\%$ for uniform sampling.
  Its mean PW remained below both baselines through the first
  300 iterations, although Hamming-based sampling subsequently overtook
  it. But for all other models the Hamming weight sampling shows better results in general.
  These results demonstrate that a model trained on simple 
  model may improve early progress per iteration on a larger interacting
  system, motivating further study of ML guidance during the initial
  optimization stage, but the Hamming weight sampling strategy is a powerful empirical choice. The final Pauli weight and wall-time comparison can be found in the Supporting Information.

\end{document}


\title{Support Information for Randomized Subsystem Descent for Fermion-to-Qubit Mapping}
\author{Gengzhi Yang}
\affiliation{\MarylandMath}
\affiliation{\MarylandQuICS}

\author{Di Wu}
\affiliation{\MarylandMath}

\author{Haizhao Yang}
\affiliation{\MarylandMath}
\affiliation{\MarylandCS}

\author{Xiaodi Wu}
\affiliation{\MarylandQuICS}
\affiliation{\MarylandCS}

\author{Ji Liu}
\affiliation{\ArgonneMathCS}
\email{ji.liu@anl.gov}

    \date{\today}

    \maketitle

    \tableofcontents

    \clearpage

\section{Supplementary tables}

  \begin{table*}[tbp]
    \centering
    \caption{Symmetry costs with Hamiltonian Pauli weight
    $\mathrm{PW}(H)$ as the objective for the molecular systems.
    Values before the arrows are the initial values, and
    values after the arrows are means $\pm$ sample standard deviation
    over 100 seeds (0-99). 
    The $\text{PW}(N), \text{PW}(S_z)$ column gives the total Pauli weight of 
    $N$ and $S_z$, and they are identical under the Clifford transformation framework.
    The last column reports the mean $\pm$ sample standard deviation
    of the maximum term weight in either transformed generator.}
    \label{tab:symmetry-pw}
    \resizebox{\textwidth}{!}{%
    \begin{tabular}{llrccc}
    \toprule
    Molecule & Basis & $n$
    & $\mathrm{PW}(H_{\mathrm{JW}})\to\mathrm{PW}(H^{\prime})$
    & $\mathrm{PW}(N),\,\mathrm{PW}(S_z)$
    & $w_{\max}'$ \\
    \midrule
    $\mathrm{H}_2$ & STO-3G & 4
    & $32\to 26.0\pm 0.0$
    & $4\to 7.0\pm 0.0$
    & $2.0\pm 0.0$ \\
    $\mathrm{LiH}$ & STO-3G & 12
    & $3\,248\to 2\,786.8\pm 47.1$
    & $12\to 17.2\pm 2.5$
    & $3.0\pm 0.8$ \\
    $\mathrm{H}_2\mathrm{O}$ & STO-3G & 14
    & $6\,332\to 5\,213.6\pm 51.2$
    & $14\to 25.3\pm 2.4$
    & $3.8\pm 0.4$ \\
    $\mathrm{CH}_4$ & STO-3G & 18
    & $42\,476\to 35\,151.0\pm 310.8$
    & $18\to 37.3\pm 3.9$
    & $4.0\pm 0.1$ \\
    $\mathrm{O}_2$ & STO-3G & 20
    & $16\,904\to 12\,721.0\pm 168.8$
    & $20\to 37.0\pm 4.0$
    & $4.0\pm 0.6$ \\
    $\mathrm{NaF}$ & STO-3G & 28
    & $247\,264\to 176\,123.4\pm 4\,734.2$
    & $28\to 72.3\pm 7.6$
    & $5.7\pm 1.0$ \\
    $\mathrm{CO}_2$ & STO-3G & 30
    & $173\,324\to 121\,572.9\pm 2\,867.9$
    & $30\to 69.1\pm 8.8$
    & $5.4\pm 0.7$ \\
    \midrule
    $\mathrm{H}_2$ & 6-31G & 8
    & $728\to 664.0\pm 0.0$
    & $8\to 15.0\pm 0.0$
    & $3.0\pm 0.0$ \\
    $\mathrm{LiH}$ & 6-31G & 22
    & $73\,872\to 55\,704.3\pm 1\,145.8$
    & $22\to 46.0\pm 3.8$
    & $5.2\pm 0.9$ \\
    $\mathrm{H}_2\mathrm{O}$ & 6-31G & 26
    & $123\,060\to 86\,124.0\pm 1\,912.2$
    & $26\to 53.9\pm 5.3$
    & $5.0\pm 0.7$ \\
    $\mathrm{CH}_4$ & 6-31G & 34
    & $899\,480\to 593\,106.4\pm 3\,677.9$
    & $34\to 71.7\pm 6.5$
    & $5.2\pm 0.4$ \\
    $\mathrm{O}_2$ & 6-31G & 36
    & $284\,280\to 174\,573.4\pm 4\,066.1$
    & $36\to 85.7\pm 7.7$
    & $5.9\pm 0.8$ \\
    $\mathrm{NaF}$ & 6-31G & 44
    & $2\,347\,244\to 1\,408\,779.2\pm 4\,480.3$
    & $44\to 127.1\pm 12.0$
    & $6.1\pm 0.4$ \\
    $\mathrm{CO}_2$ & 6-31G & 54
    & $3\,234\,356\to 1\,755\,406.6\pm 5\,888.7$
    & $54\to 149.8\pm 13.6$
    & $6.5\pm 0.5$ \\
    \bottomrule
    \end{tabular}%
    }
  \end{table*}

  \begin{table*}[tbp]
    \centering
    \caption{Same as Table~\ref{tab:symmetry-pw}, but under the weighted Pauli weight.}
    \label{tab:symmetry-wpw}
    \resizebox{\textwidth}{!}{%
    \begin{tabular}{llrccc}
    \toprule
    Molecule & Basis & $n$
    & $\mathrm{wPW}(H_{\mathrm{JW}})\to\mathrm{wPW}(H^{\prime})$
    & $\mathrm{PW}(N),\,\mathrm{PW}(S_z)$
    & $w_{\max}'$ \\
    \midrule
    $\mathrm{H}_2$ & STO-3G & 4
    & $3.355\to 3.192\pm 0.000$
    & $4\to 7.0\pm 0.0$
    & $2.0\pm 0.0$ \\
    $\mathrm{LiH}$ & STO-3G & 12
    & $28.882\to 26.087\pm 0.170$
    & $12\to 12.0\pm 0.0$
    & $1.0\pm 0.0$ \\
    $\mathrm{H}_2\mathrm{O}$ & STO-3G & 14
    & $158.681\to 135.610\pm 0.859$
    & $14\to 14.0\pm 0.0$
    & $1.0\pm 0.0$ \\
    $\mathrm{CH}_4$ & STO-3G & 18
    & $214.488\to 182.225\pm 1.430$
    & $18\to 18.1\pm 0.5$
    & $1.1\pm 0.5$ \\
    $\mathrm{O}_2$ & STO-3G & 20
    & $345.992\to 304.598\pm 2.942$
    & $20\to 20.0\pm 0.2$
    & $1.0\pm 0.2$ \\
    $\mathrm{NaF}$ & STO-3G & 28
    & $1\,224.228\to 934.061\pm 6.711$
    & $28\to 36.5\pm 0.5$
    & $3.5\pm 0.5$ \\
    $\mathrm{CO}_2$ & STO-3G & 30
    & $919.296\to 740.609\pm 16.461$
    & $30\to 38.7\pm 2.8$
    & $3.1\pm 0.7$ \\
    \midrule
    $\mathrm{H}_2$ & 6-31G & 8
    & $26.009\to 23.882\pm 0.000$
    & $8\to 8.0\pm 0.0$
    & $1.0\pm 0.0$ \\
    $\mathrm{LiH}$ & 6-31G & 22
    & $159.928\to 125.235\pm 1.507$
    & $22\to 24.5\pm 1.1$
    & $2.7\pm 0.7$ \\
    $\mathrm{H}_2\mathrm{O}$ & 6-31G & 26
    & $828.321\to 579.933\pm 10.279$
    & $26\to 33.5\pm 2.5$
    & $4.2\pm 0.5$ \\
    $\mathrm{CH}_4$ & 6-31G & 34
    & $1\,370.501\to 970.087\pm 15.863$
    & $34\to 52.1\pm 1.9$
    & $5.1\pm 0.5$ \\
    $\mathrm{O}_2$ & 6-31G & 36
    & $1\,906.171\to 1\,286.646\pm 17.884$
    & $36\to 54.9\pm 2.7$
    & $4.0\pm 0.6$ \\
    $\mathrm{NaF}$ & 6-31G & 44
    & $3\,257.456\to 1\,981.058\pm 32.094$
    & $44\to 71.3\pm 2.2$
    & $6.8\pm 1.2$ \\
    $\mathrm{CO}_2$ & 6-31G & 54
    & $6\,636.245\to 3\,894.042\pm 68.495$
    & $54\to 92.5\pm 1.4$
    & $6.1\pm 0.7$ \\
    \bottomrule
    \end{tabular}%
    }
  \end{table*}

\begingroup
\small
\setlength{\tabcolsep}{3.2pt}
\renewcommand{\arraystretch}{1.07}
\setlength{\LTcapwidth}{\textwidth}
\begin{longtable}{@{}lrrlrrr@{}}
\caption{RSD wall-clock timings for all systems tested.
$\bar{t}\pm s$ is the mean per-run optimizer wall time $\pm$ one sample
standard deviation. Here $n$ is the number of qubits, $M$ is the number of
Hamiltonian Pauli strings, and
``Start/obj.'' identifies the initial mapping and the objective function.
The iteration column gives the minimum--maximum number of completed iterations.}
\label{tab:rsd-runtime} \\
\toprule
Instance & $n$ & $M$ & Start/obj.
& \shortstack{Completed\\iterations}
& \shortstack{Per-run\\$\bar{t}\pm s$ (s)}
& \shortstack{Final PW/wPW\\mean $\pm$ SD} \\
\midrule
\endfirsthead
\multicolumn{7}{l}{\textit{Table \thetable{} continued.}} \\
\toprule
Instance & $n$ & $M$ & Start/obj.
& \shortstack{Completed\\iterations}
& \shortstack{Per-run\\$\bar{t}\pm s$ (s)}
& \shortstack{Final PW/wPW\\mean $\pm$ SD} \\
\midrule
\endhead
\midrule
\multicolumn{7}{r}{\textit{Continued on next page}} \\
\endfoot
\bottomrule
\endlastfoot
\multicolumn{7}{l}{\textbf{1D hopping: 20 sites, range $r=1,\ldots,19$}} \\*
$r=1$ & $20$ & $38$ & JW/PW & $30\,000$ & $11\,849 \pm 248$ & $56.3 \pm 0.5$ \\*
 &  &  & BK/PW & $30\,000$ & $11\,494 \pm 286$ & $56.4 \pm 0.5$ \\
$r=2$ & $20$ & $74$ & JW/PW & $30\,000$ & $13\,931 \pm 584$ & $178.0 \pm 0.0$ \\*
 &  &  & BK/PW & $30\,000$ & $11\,810 \pm 2\,018$ & $179.3 \pm 1.1$ \\
$r=3$ & $20$ & $108$ & JW/PW & $30\,000$ & $15\,123 \pm 1\,420$ & $310.0 \pm 0.0$ \\*
 &  &  & BK/PW & $30\,000$ & $11\,878 \pm 3\,324$ & $311.0 \pm 1.0$ \\
$r=4$ & $20$ & $140$ & JW/PW & $30\,000$ & $7\,679 \pm 507$ & $461.2 \pm 0.9$ \\*
 &  &  & BK/PW & $30\,000$ & $7\,990 \pm 705$ & $462.2 \pm 1.6$ \\
$r=5$ & $20$ & $170$ & JW/PW & $30\,000$ & $17\,149 \pm 2\,827$ & $600.9 \pm 2.2$ \\*
 &  &  & BK/PW & $30\,000$ & $10\,128 \pm 2\,170$ & $603.2 \pm 2.5$ \\
$r=6$ & $20$ & $198$ & JW/PW & $30\,000$ & $9\,333 \pm 350$ & $740.5 \pm 4.5$ \\*
 &  &  & BK/PW & $30\,000$ & $9\,359 \pm 349$ & $741.7 \pm 3.9$ \\
$r=7$ & $20$ & $224$ & JW/PW & $30\,000$ & $10\,169 \pm 1\,655$ & $877.1 \pm 5.1$ \\*
 &  &  & BK/PW & $30\,000$ & $10\,034 \pm 2\,325$ & $879.1 \pm 4.6$ \\
$r=8$ & $20$ & $248$ & JW/PW & $30\,000$ & $8\,302 \pm 316$ & $1\,011.1 \pm 5.7$ \\*
 &  &  & BK/PW & $30\,000$ & $8\,147 \pm 398$ & $1\,011.7 \pm 4.5$ \\
$r=9$ & $20$ & $270$ & JW/PW & $30\,000$ & $19\,393 \pm 13\,756$ & $1\,139.6 \pm 5.3$ \\*
 &  &  & BK/PW & $30\,000$ & $6\,458 \pm 465$ & $1\,141.0 \pm 6.9$ \\
$r=10$ & $20$ & $290$ & JW/PW & $30\,000$ & $8\,175 \pm 1\,255$ & $1\,262.0 \pm 4.6$ \\*
 &  &  & BK/PW & $30\,000$ & $8\,346 \pm 1\,856$ & $1\,259.3 \pm 5.7$ \\
$r=11$ & $20$ & $308$ & JW/PW & $30\,000$ & $12\,408 \pm 918$ & $1\,367.9 \pm 6.2$ \\*
 &  &  & BK/PW & $30\,000$ & $11\,036 \pm 814$ & $1\,364.1 \pm 6.1$ \\
$r=12$ & $20$ & $324$ & JW/PW & $30\,000$ & $14\,269 \pm 777$ & $1\,454.8 \pm 6.5$ \\*
 &  &  & BK/PW & $30\,000$ & $11\,693 \pm 3\,392$ & $1\,453.5 \pm 6.5$ \\
$r=13$ & $20$ & $338$ & JW/PW & $30\,000$ & $8\,572 \pm 826$ & $1\,530.3 \pm 6.3$ \\*
 &  &  & BK/PW & $30\,000$ & $8\,674 \pm 597$ & $1\,528.6 \pm 5.7$ \\
$r=14$ & $20$ & $350$ & JW/PW & $30\,000$ & $7\,257 \pm 186$ & $1\,595.5 \pm 6.2$ \\*
 &  &  & BK/PW & $30\,000$ & $7\,149 \pm 172$ & $1\,594.8 \pm 5.8$ \\
$r=15$ & $20$ & $360$ & JW/PW & $30\,000$ & $8\,883 \pm 1\,559$ & $1\,650.7 \pm 5.3$ \\*
 &  &  & BK/PW & $30\,000$ & $8\,721 \pm 2\,006$ & $1\,648.3 \pm 5.1$ \\
$r=16$ & $20$ & $368$ & JW/PW & $30\,000$ & $8\,011 \pm 1\,713$ & $1\,693.8 \pm 4.1$ \\*
 &  &  & BK/PW & $30\,000$ & $7\,703 \pm 1\,592$ & $1\,691.5 \pm 4.5$ \\
$r=17$ & $20$ & $374$ & JW/PW & $30\,000$ & $8\,537 \pm 368$ & $1\,727.2 \pm 4.4$ \\*
 &  &  & BK/PW & $30\,000$ & $8\,298 \pm 280$ & $1\,723.2 \pm 3.7$ \\
$r=18$ & $20$ & $378$ & JW/PW & $30\,000$ & $8\,550 \pm 533$ & $1\,746.3 \pm 3.7$ \\*
 &  &  & BK/PW & $30\,000$ & $8\,513 \pm 345$ & $1\,744.7 \pm 4.1$ \\
$r=19$ & $20$ & $380$ & JW/PW & $30\,000$ & $11\,235 \pm 2\,212$ & $1\,757.0 \pm 4.1$ \\*
 &  &  & BK/PW & $30\,000$ & $6\,903 \pm 955$ & $1\,756.6 \pm 3.5$ \\
\midrule
\multicolumn{7}{l}{\textbf{1D all-to-all hopping, system size $N=2,\ldots,20$}} \\*
$N=2$ & $2$ & $2$ & JW/PW & $2$ & $0.000802 \pm 0.000985$ & $2.0 \pm 0.0$ \\*
 &  &  & BK/PW & $2$ & $0.000431 \pm 0.000060$ & $2.0 \pm 0.0$ \\
$N=3$ & $3$ & $6$ & JW/PW & $2$ & $0.000890 \pm 0.000480$ & $10.0 \pm 0.0$ \\*
 &  &  & BK/PW & $2$ & $0.000609 \pm 0.000060$ & $10.0 \pm 0.0$ \\
$N=4$ & $4$ & $12$ & JW/PW & $3$ & $0.004383 \pm 0.002392$ & $24.0 \pm 0.0$ \\*
 &  &  & BK/PW & $3$ & $0.002887 \pm 0.000305$ & $24.0 \pm 0.0$ \\
$N=5$ & $5$ & $20$ & JW/PW & $3$ & $0.0211 \pm 0.0127$ & $48.0 \pm 0.0$ \\*
 &  &  & BK/PW & $3$ & $0.0130 \pm 0.0030$ & $48.0 \pm 0.0$ \\
$N=6$ & $6$ & $30$ & JW/PW & $5$ & $0.159 \pm 0.089$ & $79.0 \pm 0.0$ \\*
 &  &  & BK/PW & $4$ & $0.108 \pm 0.054$ & $79.0 \pm 0.0$ \\
$N=7$ & $7$ & $42$ & JW/PW & $5$ & $0.349 \pm 0.031$ & $122.0 \pm 0.0$ \\*
 &  &  & BK/PW & $5$ & $0.351 \pm 0.044$ & $122.0 \pm 0.0$ \\
$N=8$ & $8$ & $56$ & JW/PW & $7$ & $1.89 \pm 0.61$ & $173.0 \pm 0.0$ \\*
 &  &  & BK/PW & $5$ & $1.10 \pm 0.24$ & $173.0 \pm 0.0$ \\
$N=9$ & $9$ & $72$ & JW/PW & $16$--$23$ & $6.26 \pm 1.17$ & $234.9 \pm 2.9$ \\*
 &  &  & BK/PW & $14$--$20$ & $5.67 \pm 1.16$ & $233.7 \pm 1.1$ \\
$N=10$ & $10$ & $90$ & JW/PW & $52$--$64$ & $19.2 \pm 10.3$ & $312.0 \pm 1.6$ \\*
 &  &  & BK/PW & $50$--$58$ & $10.6 \pm 1.2$ & $311.5 \pm 1.0$ \\
$N=11$ & $11$ & $110$ & JW/PW & $172$--$189$ & $84.7 \pm 6.7$ & $399.4 \pm 2.6$ \\*
 &  &  & BK/PW & $172$--$188$ & $81.4 \pm 8.6$ & $398.8 \pm 1.6$ \\
$N=12$ & $12$ & $132$ & JW/PW & $504$--$539$ & $233.5 \pm 112.6$ & $499.1 \pm 3.4$ \\*
 &  &  & BK/PW & $502$--$518$ & $180.2 \pm 30.7$ & $495.8 \pm 2.8$ \\
$N=13$ & $13$ & $156$ & JW/PW & $1\,300$--$1\,356$ & $599.6 \pm 547.0$ & $611.3 \pm 2.9$ \\*
 &  &  & BK/PW & $1\,296$--$1\,327$ & $265.7 \pm 17.9$ & $610.1 \pm 2.6$ \\
$N=14$ & $14$ & $182$ & JW/PW & $3\,018$--$3\,051$ & $936.9 \pm 68.4$ & $733.5 \pm 2.5$ \\*
 &  &  & BK/PW & $3\,014$--$3\,051$ & $955.0 \pm 53.9$ & $732.9 \pm 1.8$ \\
$N=15$ & $15$ & $210$ & JW/PW & $6\,450$--$6\,574$ & $2\,512 \pm 195$ & $872.5 \pm 4.0$ \\*
 &  &  & BK/PW & $6\,449$--$6\,581$ & $2\,129 \pm 528$ & $868.7 \pm 3.7$ \\
$N=16$ & $16$ & $240$ & JW/PW & $12\,886$--$13\,065$ & $4\,277 \pm 942$ & $1\,018.0 \pm 3.5$ \\*
 &  &  & BK/PW & $12\,885$--$12\,975$ & $2\,662 \pm 146$ & $1\,015.6 \pm 2.3$ \\
$N=17$ & $17$ & $272$ & JW/PW & $24\,333$--$24\,448$ & $14\,600 \pm 6\,006$ & $1\,177.2 \pm 5.4$ \\*
 &  &  & BK/PW & $24\,328$--$24\,427$ & $12\,120 \pm 4\,390$ & $1\,176.8 \pm 6.3$ \\
$N=18$ & $18$ & $306$ & JW/PW & $30\,000$ & $13\,262 \pm 4\,563$ & $1\,353.2 \pm 5.6$ \\*
 &  &  & BK/PW & $30\,000$ & $12\,152 \pm 2\,664$ & $1\,349.6 \pm 8.8$ \\
$N=19$ & $19$ & $342$ & JW/PW & $30\,000$ & $11\,396 \pm 2\,622$ & $1\,546.7 \pm 5.1$ \\*
 &  &  & BK/PW & $30\,000$ & $12\,832 \pm 1\,674$ & $1\,545.8 \pm 4.4$ \\
$N=20$ & $20$ & $380$ & JW/PW & $30\,000$ & $12\,272 \pm 7\,683$ & $1\,757.0 \pm 4.1$ \\*
 &  &  & BK/PW & $30\,000$ & $9\,497 \pm 736$ & $1\,756.6 \pm 3.5$ \\
\midrule
\multicolumn{7}{l}{\textbf{2D nearest-neighbor hopping, grid $L\times L$, $L=2,\ldots,16$}} \\*
Hopping, $L=2$ & $4$ & $8$ & JW/PW & $3$ & $0.003787 \pm 0.002041$ & $12.0 \pm 0.0$ \\*
 &  &  & BK/PW & $3$ & $0.002687 \pm 0.000157$ & $12.0 \pm 0.0$ \\
Hopping, $L=3$ & $9$ & $24$ & JW/PW & $13$--$17$ & $4.14 \pm 1.25$ & $45.9 \pm 0.5$ \\*
 &  &  & BK/PW & $14$--$18$ & $4.39 \pm 0.99$ & $44.8 \pm 1.0$ \\
Hopping, $L=4$ & $16$ & $48$ & JW/PW & $12\,889$--$13\,025$ & $4\,082 \pm 1\,099$ & $102.8 \pm 2.2$ \\*
 &  &  & BK/PW & $12\,883$--$13\,254$ & $4\,051 \pm 1\,105$ & $107.1 \pm 1.9$ \\
Hopping, $L=5$ & $25$ & $80$ & JW/PW & $30\,000$ & $15\,882 \pm 5\,591$ & $184.1 \pm 1.7$ \\*
 &  &  & BK/PW & $30\,000$ & $11\,283 \pm 914$ & $184.3 \pm 1.9$ \\
Hopping, $L=6$ & $36$ & $120$ & JW/PW & $30\,000$ & $15\,540 \pm 2\,559$ & $293.2 \pm 3.0$ \\*
 &  &  & BK/PW & $30\,000$ & $12\,157 \pm 2\,096$ & $296.0 \pm 4.0$ \\
Hopping, $L=7$ & $49$ & $168$ & JW/PW & $30\,000$ & $12\,125 \pm 359$ & $434.7 \pm 3.1$ \\*
 &  &  & BK/PW & $30\,000$ & $10\,541 \pm 594$ & $435.9 \pm 4.8$ \\
Hopping, $L=8$ & $64$ & $224$ & JW/PW & $30\,000$ & $14\,297 \pm 521$ & $613.4 \pm 4.7$ \\*
 &  &  & BK/PW & $30\,000$ & $10\,584 \pm 3\,716$ & $619.7 \pm 4.6$ \\
Hopping, $L=9$ & $81$ & $288$ & JW/PW & $30\,000$ & $7\,983 \pm 422$ & $819.6 \pm 14.0$ \\*
 &  &  & BK/PW & $30\,000$ & $7\,949 \pm 683$ & $821.7 \pm 16.0$ \\
Hopping, $L=10$ & $100$ & $360$ & JW/PW & $30\,000$ & $17\,088 \pm 8\,703$ & $1\,070.4 \pm 24.1$ \\*
 &  &  & BK/PW & $30\,000$ & $8\,493 \pm 492$ & $1\,071.2 \pm 21.8$ \\
Hopping, $L=11$ & $121$ & $440$ & JW/PW & $30\,000$ & $7\,033 \pm 930$ & $1\,391.0 \pm 75.1$ \\*
 &  &  & BK/PW & $30\,000$ & $7\,436 \pm 520$ & $1\,381.8 \pm 69.7$ \\
Hopping, $L=12$ & $144$ & $528$ & JW/PW & $30\,000$ & $16\,229 \pm 6\,739$ & $1\,878.7 \pm 97.0$ \\*
 &  &  & BK/PW & $30\,000$ & $6\,466 \pm 1\,492$ & $1\,824.0 \pm 97.3$ \\
Hopping, $L=13$ & $169$ & $624$ & JW/PW & $30\,000$ & $8\,936 \pm 759$ & $2\,460.7 \pm 98.1$ \\*
 &  &  & BK/PW & $30\,000$ & $9\,885 \pm 1\,656$ & $2\,560.3 \pm 80.7$ \\
Hopping, $L=14$ & $196$ & $728$ & JW/PW & $30\,000$ & $9\,135 \pm 1\,406$ & $3\,074.9 \pm 69.0$ \\*
 &  &  & BK/PW & $30\,000$ & $10\,517 \pm 1\,673$ & $3\,044.7 \pm 71.8$ \\
Hopping, $L=15$ & $225$ & $840$ & JW/PW & $30\,000$ & $9\,593 \pm 329$ & $3\,751.6 \pm 77.3$ \\*
 &  &  & BK/PW & $30\,000$ & $8\,681 \pm 829$ & $3\,800.6 \pm 54.4$ \\
Hopping, $L=16$ & $256$ & $960$ & JW/PW & $30\,000$ & $19\,267 \pm 4\,210$ & $4\,451.8 \pm 67.3$ \\*
 &  &  & BK/PW & $30\,000$ & $7\,570 \pm 1\,259$ & $4\,184.0 \pm 60.0$ \\
\midrule
\multicolumn{7}{l}{\textbf{2D Hubbard model, grid $L\times L$, $L=2,\ldots,16$}} \\*
Hubbard, $L=2$ & $8$ & $29$ & JW/PW & $4$ & $1.57 \pm 0.98$ & $56.0 \pm 0.0$ \\*
 &  &  & BK/PW & $8$ & $4.70 \pm 2.15$ & $56.0 \pm 0.0$ \\
Hubbard, $L=3$ & $18$ & $76$ & JW/PW & $30\,000$ & $12\,558 \pm 574$ & $180.9 \pm 1.5$ \\*
 &  &  & BK/PW & $30\,000$ & $11\,914 \pm 405$ & $183.5 \pm 7.3$ \\
Hubbard, $L=4$ & $32$ & $145$ & JW/PW & $30\,000$ & $12\,608 \pm 431$ & $387.4 \pm 5.8$ \\*
 &  &  & BK/PW & $30\,000$ & $11\,764 \pm 546$ & $411.6 \pm 27.2$ \\
Hubbard, $L=5$ & $50$ & $236$ & JW/PW & $30\,000$ & $11\,326 \pm 3\,278$ & $686.7 \pm 18.7$ \\*
 &  &  & BK/PW & $30\,000$ & $12\,925 \pm 3\,723$ & $705.0 \pm 32.8$ \\
Hubbard, $L=6$ & $72$ & $349$ & JW/PW & $30\,000$ & $10\,657 \pm 1\,322$ & $1\,095.6 \pm 40.6$ \\*
 &  &  & BK/PW & $30\,000$ & $8\,130 \pm 1\,300$ & $1\,167.8 \pm 74.9$ \\
Hubbard, $L=7$ & $98$ & $484$ & JW/PW & $30\,000$ & $12\,212 \pm 3\,540$ & $1\,707.7 \pm 88.6$ \\*
 &  &  & BK/PW & $30\,000$ & $7\,998 \pm 1\,439$ & $1\,831.0 \pm 59.0$ \\
Hubbard, $L=8$ & $128$ & $641$ & JW/PW & $30\,000$ & $7\,376 \pm 461$ & $2\,534.4 \pm 71.1$ \\*
 &  &  & BK/PW & $30\,000$ & $7\,954 \pm 400$ & $2\,638.3 \pm 35.7$ \\
Hubbard, $L=9$ & $162$ & $820$ & JW/PW & $30\,000$ & $8\,404 \pm 508$ & $3\,466.5 \pm 50.8$ \\*
 &  &  & BK/PW & $30\,000$ & $7\,847 \pm 605$ & $3\,558.4 \pm 49.4$ \\
Hubbard, $L=10$ & $200$ & $1\,021$ & JW/PW & $30\,000$ & $7\,112 \pm 439$ & $4\,538.1 \pm 55.2$ \\*
 &  &  & BK/PW & $30\,000$ & $7\,474 \pm 514$ & $4\,573.8 \pm 44.5$ \\
Hubbard, $L=11$ & $242$ & $1\,244$ & JW/PW & $30\,000$ & $8\,935 \pm 2\,582$ & $5\,730.1 \pm 52.4$ \\*
 &  &  & BK/PW & $30\,000$ & $9\,107 \pm 2\,369$ & $5\,772.4 \pm 53.0$ \\
Hubbard, $L=12$ & $288$ & $1\,489$ & JW/PW & $30\,000$ & $21\,051 \pm 6\,078$ & $7\,087.3 \pm 63.5$ \\*
 &  &  & BK/PW & $30\,000$ & $9\,998 \pm 1\,788$ & $6\,984.5 \pm 39.7$ \\
Hubbard, $L=13$ & $338$ & $1\,756$ & JW/PW & $30\,000$ & $17\,049 \pm 963$ & $8\,622.1 \pm 77.6$ \\*
 &  &  & BK/PW & $30\,000$ & $12\,583 \pm 886$ & $8\,592.2 \pm 57.0$ \\
Hubbard, $L=14$ & $392$ & $2\,045$ & JW/PW & $30\,000$ & $14\,631 \pm 517$ & $10\,294.4 \pm 96.8$ \\*
 &  &  & BK/PW & $30\,000$ & $15\,025 \pm 238$ & $10\,019.0 \pm 74.3$ \\
Hubbard, $L=15$ & $450$ & $2\,356$ & JW/PW & $30\,000$ & $17\,045 \pm 561$ & $12\,263.3 \pm 145.1$ \\*
 &  &  & BK/PW & $30\,000$ & $13\,969 \pm 4\,062$ & $11\,821.6 \pm 64.4$ \\
Hubbard, $L=16$ & $512$ & $2\,689$ & JW/PW & $30\,000$ & $13\,923 \pm 3\,063$ & $14\,485.5 \pm 201.9$ \\*
 &  &  & BK/PW & $30\,000$ & $13\,411 \pm 2\,114$ & $13\,254.3 \pm 55.4$ \\
\midrule
\multicolumn{7}{l}{\textbf{Molecular systems: STO-3G}} \\*
$\mathrm{H}_2$ & $4$ & $15$ & JW/PW & $2$ & $0.005591 \pm 0.010213$ & $26.0 \pm 0.0$ \\*
 &  &  & JW/wPW & $2$ & $0.002490 \pm 0.002535$ & $3.192 \pm 0.000$ \\
$\mathrm{LiH}$ & $12$ & $631$ & JW/PW & $498$--$517$ & $95.6 \pm 2.8$ & $2\,786.8 \pm 47.1$ \\*
 &  &  & JW/wPW & $499$--$527$ & $87.8 \pm 2.6$ & $26.087 \pm 0.170$ \\
$\mathrm{H}_2\mathrm{O}$ & $14$ & $1\,086$ & JW/PW & $2\,000$ & $342.0 \pm 3.8$ & $5\,213.6 \pm 51.2$ \\*
 &  &  & JW/wPW & $2\,000$ & $339.2 \pm 10.5$ & $135.610 \pm 0.859$ \\
$\mathrm{CH}_4$ & $18$ & $5\,956$ & JW/PW & $2\,000$ & $399.2 \pm 43.1$ & $35\,151.0 \pm 310.8$ \\*
 &  &  & JW/wPW & $2\,000$ & $450.6 \pm 30.2$ & $182.225 \pm 1.430$ \\
$\mathrm{O}_2$ & $20$ & $2\,239$ & JW/PW & $2\,000$ & $437.6 \pm 23.6$ & $12\,721.0 \pm 168.8$ \\*
 &  &  & JW/wPW & $2\,000$ & $457.5 \pm 36.5$ & $304.598 \pm 2.942$ \\
$\mathrm{NaF}$ & $28$ & $24\,275$ & JW/PW & $2\,000$ & $442.2 \pm 35.2$ & $176\,123.4 \pm 4\,734.2$ \\*
 &  &  & JW/wPW & $2\,000$ & $351.6 \pm 10.7$ & $934.061 \pm 6.711$ \\
$\mathrm{CO}_2$ & $30$ & $16\,170$ & JW/PW & $2\,000$ & $339.1 \pm 2.3$ & $121\,572.9 \pm 2\,867.9$ \\*
 &  &  & JW/wPW & $2\,000$ & $605.4 \pm 38.3$ & $740.609 \pm 16.461$ \\
\midrule
\multicolumn{7}{l}{\textbf{Molecular systems: 6-31G}} \\*
$\mathrm{H}_2$ & $8$ & $185$ & JW/PW & $5$ & $1.33 \pm 0.14$ & $664.0 \pm 0.0$ \\*
 &  &  & JW/wPW & $4$ & $0.969 \pm 0.097$ & $23.882 \pm 0.000$ \\
$\mathrm{LiH}$ & $22$ & $8\,758$ & JW/PW & $2\,000$ & $354.4 \pm 24.7$ & $55\,704.3 \pm 1\,145.8$ \\*
 &  &  & JW/wPW & $2\,000$ & $345.0 \pm 26.4$ & $125.235 \pm 1.507$ \\
$\mathrm{H}_2\mathrm{O}$ & $26$ & $12\,732$ & JW/PW & $2\,000$ & $330.36 \pm 0.96$ & $86\,124.0 \pm 1\,912.2$ \\*
 &  &  & JW/wPW & $2\,000$ & $356.9 \pm 14.0$ & $579.933 \pm 10.279$ \\
$\mathrm{CH}_4$ & $34$ & $75\,444$ & JW/PW & $2\,000$ & $402.8 \pm 41.2$ & $593\,106.4 \pm 3\,677.9$ \\*
 &  &  & JW/wPW & $2\,000$ & $914.1 \pm 118.6$ & $970.087 \pm 15.863$ \\
$\mathrm{O}_2$ & $36$ & $22\,543$ & JW/PW & $2\,000$ & $549.6 \pm 59.0$ & $174\,573.4 \pm 4\,066.1$ \\*
 &  &  & JW/wPW & $2\,000$ & $345.9 \pm 7.4$ & $1\,286.646 \pm 17.884$ \\
$\mathrm{NaF}$ & $44$ & $157\,607$ & JW/PW & $2\,000$ & $639.3 \pm 160.0$ & $1\,408\,779.2 \pm 4\,480.3$ \\*
 &  &  & JW/wPW & $2\,000$ & $485.6 \pm 40.4$ & $1\,981.058 \pm 32.094$ \\
$\mathrm{CO}_2$ & $54$ & $182\,270$ & JW/PW & $2\,000$ & $496.7 \pm 34.7$ & $1\,755\,406.6 \pm 5\,888.7$ \\*
 &  &  & JW/wPW & $2\,000$ & $446.3 \pm 29.2$ & $3\,894.042 \pm 68.495$ \\
\end{longtable}
\endgroup

\begingroup
\small
\setlength{\tabcolsep}{3pt}
\renewcommand{\arraystretch}{1.10}
\setlength{\LTcapwidth}{\textwidth}
\begin{longtable}{@{}lrlrrrr@{}}
\caption{Training and testing costs of the ML-guided RSD experiment.
Panel (a) summarizes five models trained with seeds 0--4. Panel (b) compares 5000 RSD
iterations from Jordan--Wigner with width being $8$ and depth being $4$.
Timings use 50 executions per method and system. Selection includes prediction, and both are included in RSD time. Bold entries identify the smallest observed mean final PW in each system.}\label{tab:sampling-costs} \\
\toprule
\multicolumn{7}{l}{\textbf{(a) Training cost per model}} \\*
\multicolumn{4}{l}{Training-data generation (13,500 examples)} & \multicolumn{3}{r}{$202.2\pm 87.8$\,s} \\*
\multicolumn{4}{l}{XGBoost fitting} & \multicolumn{3}{r}{$1.923\pm 0.964$\,s} \\*
\multicolumn{4}{l}{Total training: generation + fitting} & \multicolumn{3}{r}{$204.1\pm 88.4$\,s} \\*
\multicolumn{4}{l}{Allocated training cost per ML test run (total/70)} & \multicolumn{3}{r}{$2.916\pm 1.264$\,s} \\
\midrule
\multicolumn{7}{l}{\textbf{(b) Testing: cost per RSD run and final PW}} \\*
System & $n$ & Sampling
& \shortstack{RSD time\\(s)} & \shortstack{Selection\\(s)}
& \shortstack{Prediction\\(s)} & \shortstack{Final PW\\mean $\pm$ SD} \\
\midrule
\endfirsthead
\multicolumn{7}{l}{\textit{Table \thetable{} continued.}} \\
\toprule
System & $n$ & Sampling
& \shortstack{RSD time\\(s)} & \shortstack{Selection\\(s)}
& \shortstack{Prediction\\(s)} & \shortstack{Final PW\\mean $\pm$ SD} \\
\midrule
\endhead
\midrule
\multicolumn{7}{r}{\textit{Continued on next page}} \\
\endfoot
\bottomrule
\endlastfoot
1D hopping ($r=20$) & 30 & Uniform & $2\,382.1\pm 1\,091.8$ & $1.34\pm 0.76$ & --- & $4\,086.8\pm 14.8$ \\*
 &  & Hamming & $2\,376.5\pm 1\,076.9$ & $1.36\pm 0.69$ & --- & $\boldsymbol{4\,084.6\pm 14.3}$ \\*
 &  & ML-guided & $2\,667.3\pm 1\,168.9$ & $119.6\pm 56.3$ & $36.1\pm 18.7$ & $4\,115.3\pm 26.2$ \\
\midrule
2D hopping ($6\times6$) & 36 & Uniform & $2\,373.3\pm 1\,088.9$ & $1.51\pm 0.83$ & --- & $294.4\pm 3.4$ \\*
 &  & Hamming & $2\,517.9\pm 972.7$ & $1.62\pm 0.63$ & --- & $\boldsymbol{293.6\pm 3.0}$ \\*
 &  & ML-guided & $2\,873.3\pm 944.3$ & $112.0\pm 42.2$ & $38.5\pm 10.9$ & $295.2\pm 2.5$ \\
\midrule
2D hopping ($8\times8$) & 64 & Uniform & $2\,754.3\pm 889.6$ & $1.67\pm 0.64$ & --- & $614.1\pm 5.5$ \\*
 &  & Hamming & $2\,851.3\pm 858.0$ & $1.79\pm 0.74$ & --- & $\boldsymbol{613.9\pm 4.9}$ \\*
 &  & ML-guided & $3\,080.3\pm 867.8$ & $121.5\pm 42.0$ & $42.1\pm 12.5$ & $618.8\pm 12.9$ \\
\midrule
2D hopping ($10\times10$) & 100 & Uniform & $2\,800.0\pm 839.8$ & $1.58\pm 0.60$ & --- & $1\,341.5\pm 42.4$ \\*
 &  & Hamming & $2\,776.3\pm 787.0$ & $1.55\pm 0.58$ & --- & $\boldsymbol{1\,217.0\pm 72.4}$ \\*
 &  & ML-guided & $3\,073.5\pm 861.9$ & $125.1\pm 39.8$ & $42.3\pm 12.3$ & $1\,349.7\pm 66.0$ \\
\midrule
2D Hubbard ($6\times6$) & 72 & Uniform & $2\,826.6\pm 820.5$ & $1.57\pm 0.55$ & --- & $1\,207.9\pm 60.7$ \\*
 &  & Hamming & $2\,863.1\pm 856.2$ & $1.58\pm 0.52$ & --- & $1\,181.4\pm 47.6$ \\*
 &  & ML-guided & $3\,014.4\pm 903.3$ & $120.7\pm 37.1$ & $40.6\pm 11.2$ & $\boldsymbol{1\,173.3\pm 66.2}$ \\
\midrule
2D Hubbard ($8\times8$) & 128 & Uniform & $2\,729.6\pm 896.1$ & $1.49\pm 0.56$ & --- & $2\,739.1\pm 33.4$ \\*
 &  & Hamming & $2\,438.4\pm 850.6$ & $1.35\pm 0.45$ & --- & $\boldsymbol{2\,711.5\pm 33.1}$ \\*
 &  & ML-guided & $2\,363.7\pm 1\,006.7$ & $106.7\pm 34.6$ & $31.0\pm 14.7$ & $2\,761.5\pm 43.8$ \\
\midrule
2D Hubbard ($10\times10$) & 200 & Uniform & $2\,111.6\pm 950.6$ & $1.00\pm 0.44$ & --- & $4\,905.2\pm 71.8$ \\*
 &  & Hamming & $1\,505.3\pm 672.7$ & $0.80\pm 0.38$ & --- & $\boldsymbol{4\,733.5\pm 77.7}$ \\*
 &  & ML-guided & $1\,198.3\pm 332.0$ & $75.0\pm 8.2$ & $13.8\pm 3.6$ & $4\,827.3\pm 59.2$ \\
\end{longtable}
\endgroup

\clearpage
\section{Supplementary figures}

\begin{figure}
    \centering
    \includegraphics[width=1.0\linewidth]{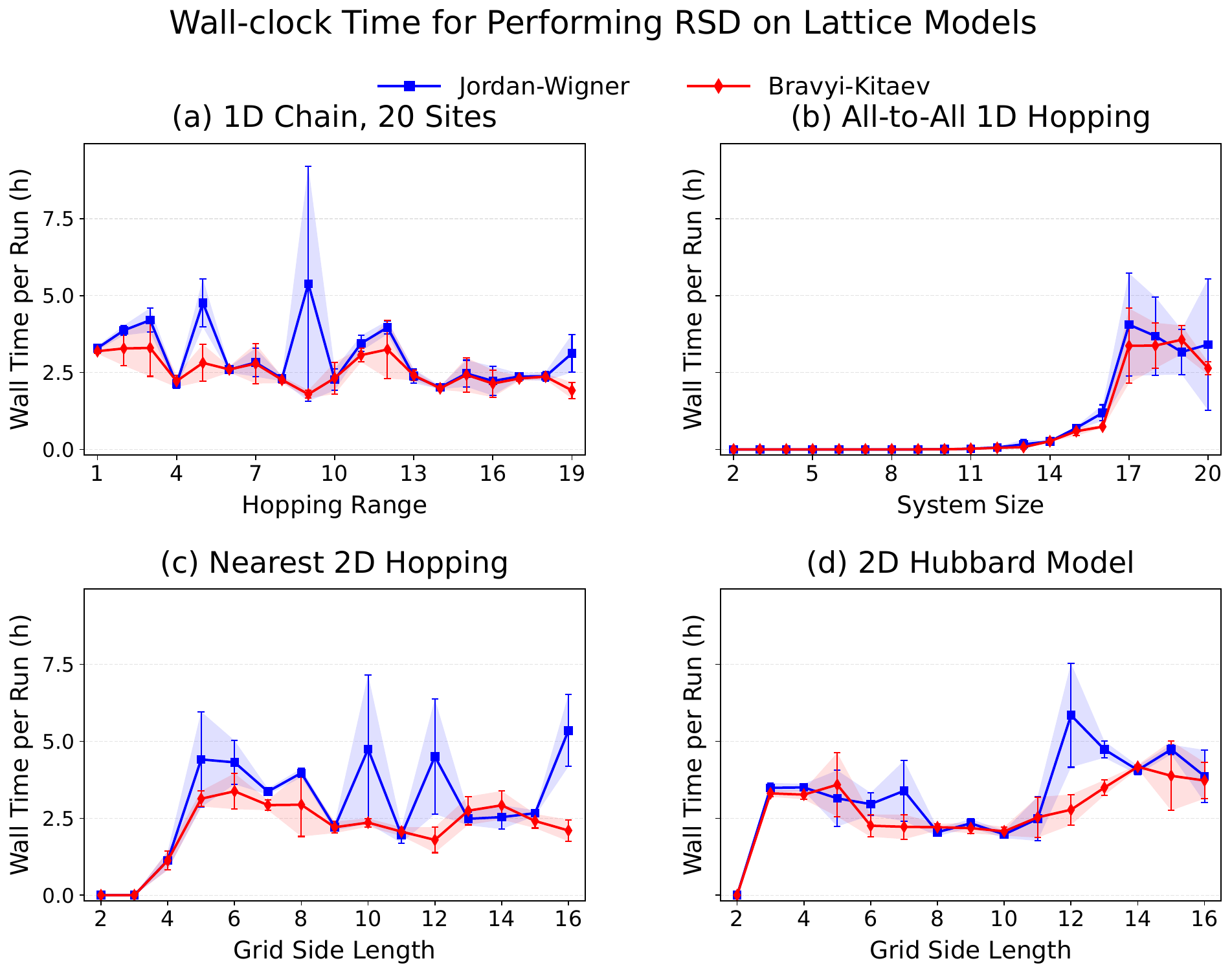}
    \caption{Wall-clock time for (a) a 20-site
  one-dimensional hopping chain with varying hopping range,
  (b) all-to-all one-dimensional hopping with varying system size,
  (c) nearest-neighbor hopping on square grids, and
  (d) the square-grid Hubbard model.
  Blue squares and red diamonds denote Jordan--Wigner and
  Bravyi--Kitaev initialization, respectively.
  Markers show mean optimizer wall time over 100 seeds (0--99), and 
  error bars and shaded bands indicate $\pm$ one sample standard
  deviation.}
\label{fig:lattice_runtime}
\end{figure}

\begin{figure}
    \centering
    \includegraphics[width=1.0\linewidth]{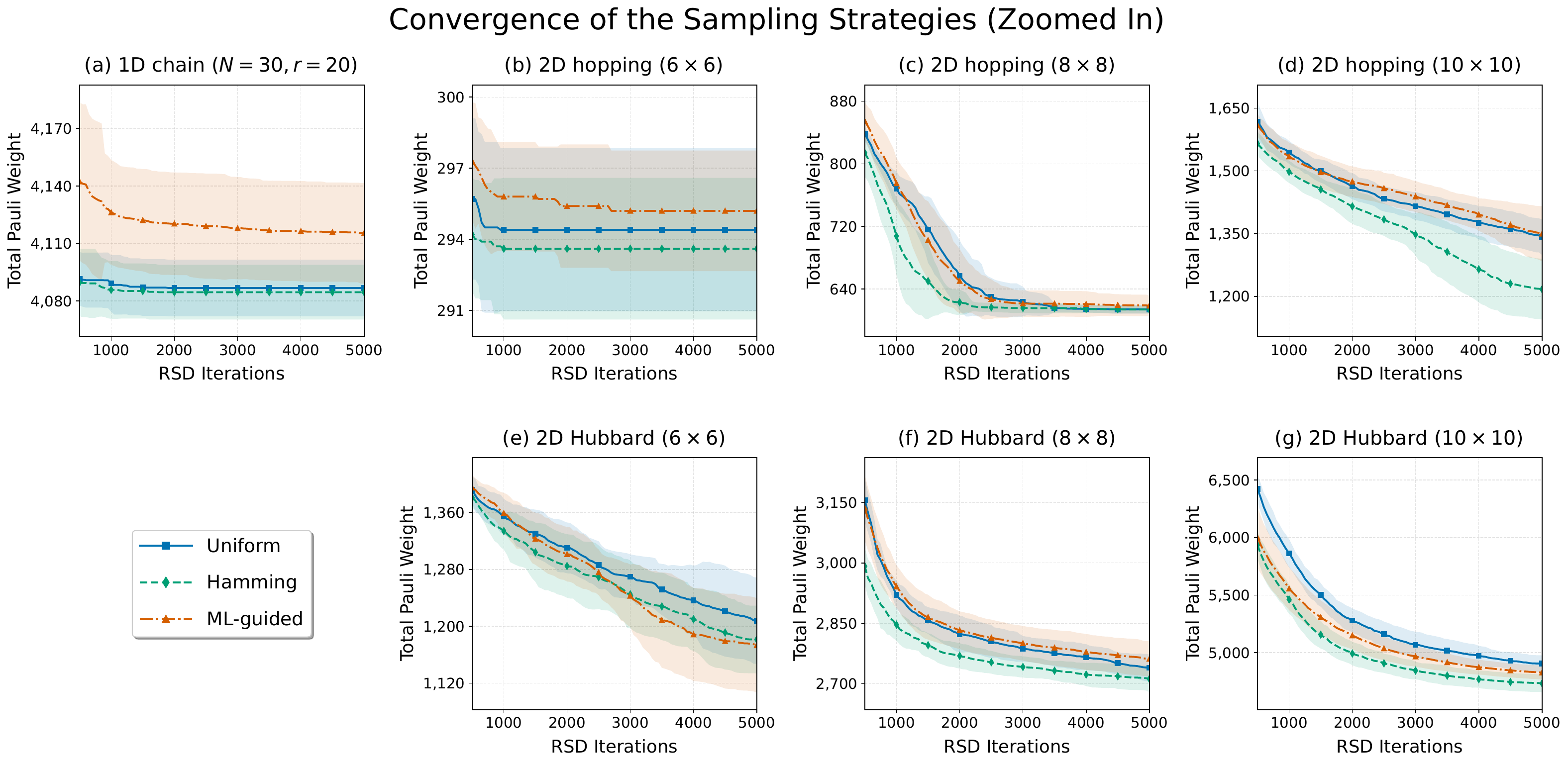}
    \caption{Convergence of uniform, Hamming-based, and ML-guided
  subsystem sampling, enlarged over iterations 500--5000.
  Markers show mean total Hamiltonian Pauli weight, and shaded bands indicate $\pm$ one sample standard
  deviation.}
    \label{fig:sampling_late}
\end{figure}

\begin{figure}
    \centering
    \includegraphics[width=1.0\linewidth]{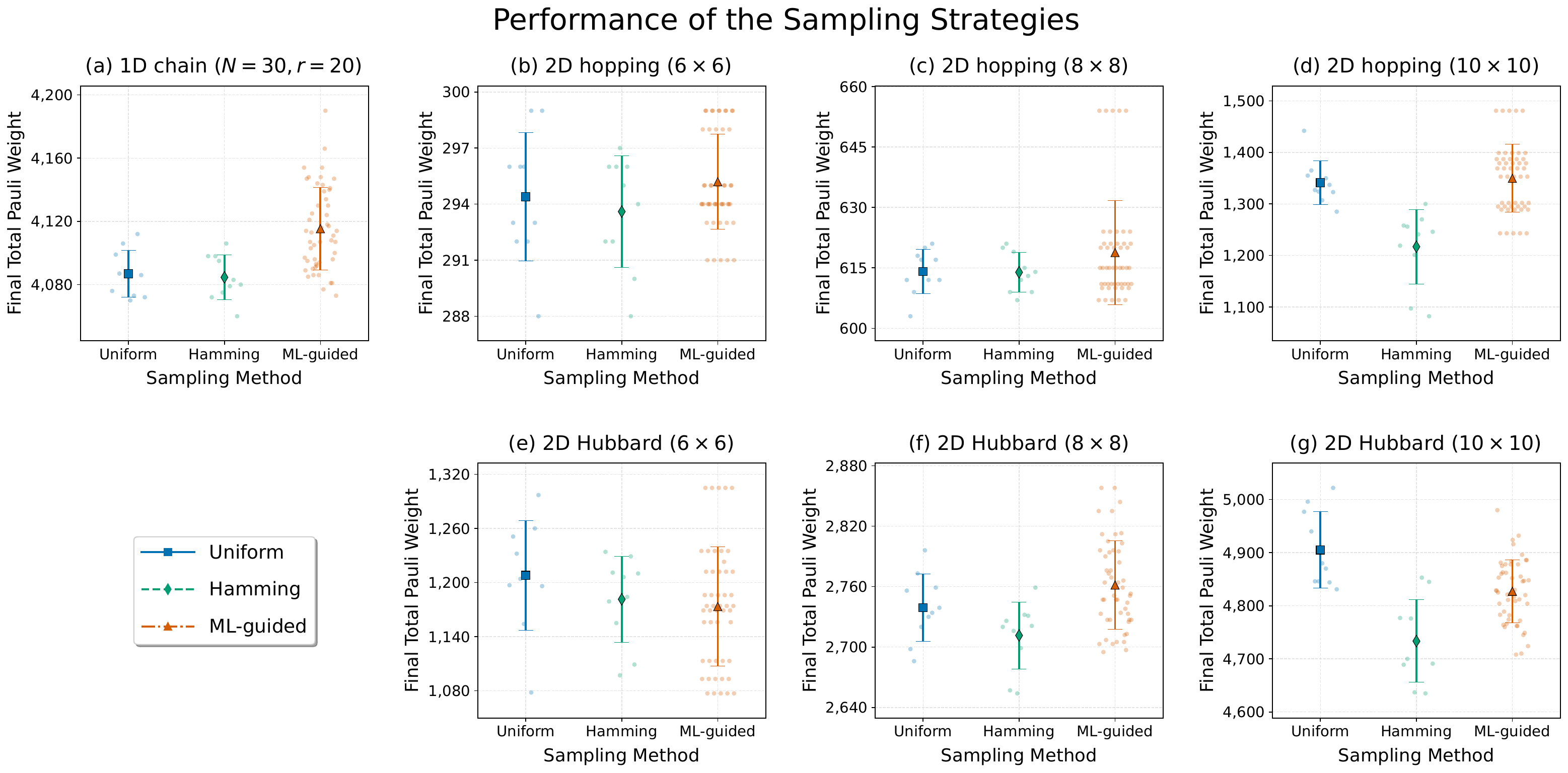}
    \caption{Final total Hamiltonian Pauli weight obtained with
  uniform, Hamming-based, and ML-guided subsystem sampling after
  5000 RSD iterations.
  Large markers and error bars show means $\pm$ one sample
  standard deviation, and faint points show individual observations.}
    \label{fig:sampling_pw}
\end{figure}